\documentclass[letterpaper,twocolumn,10pt]{article}

\usepackage{amsmath}
\usepackage{bbm}
\usepackage{amssymb}
\DeclareMathOperator*{\argmax}{arg\,max}

\usepackage{algorithm}
\usepackage{algpseudocode}
\usepackage{makecell}
\usepackage{threeparttable}
\usepackage{enumitem}
\usepackage{multirow}

\usepackage{xspace}
\newcommand{\system}{{\textsc{XSec}}\xspace}

\usepackage{tikz}
\DeclareRobustCommand*\circled[1]{\tikz[baseline=(char.base)]{ \node[shape=circle,draw,color=white,fill=black,inner sep=0.5pt] (char){#1};}}
\DeclareRobustCommand*\circledRed[1]{\tikz[baseline=(char.base)]{ \node[shape=circle,draw,color=white,fill=red,inner sep=0.5pt, font=\bfseries] (char){#1};}}

\newcommand{\shortsectionBf}[1]{\vspace{3pt}\noindent{\bf #1}}
\newcommand{\shortsectionEmph}[1]{\vspace{3pt}\noindent{\em #1}}
\def\ie{{i.e.},~}
\def\eg{{e.g.},~}

\usepackage{xcolor}
\usepackage{soul}

\usepackage{usenix}

\newcommand{\authorrowbreak}{%
\end{tabular}\par\vspace{0.5em}\begin{tabular}[t]{c}
}

\begin{document}

\pagestyle{empty}
\date{}

\title{
{
\normalsize \normalfont
Extended version of \emph{A Self-Explainable Deep Architecture for Security Applications}\\\vspace{-0.5em}
(Accepted at the Conference on Game Theory and AI for Security (GameSec), 2026)
}\\
\vspace{1em}
\Large \bf A Self-Explainable Deep Architecture for Security Applications
}

\author{
{\rm Ananth Shreekumar}\\
Purdue University
\and
{\rm Jyun-Jhu Syu}\\
Purdue University
\and
{\rm Muslum Ozgur Ozmen}\\
Arizona State University
\authorrowbreak
{\rm Dongyan Xu}\\
Purdue University
\and
{\rm Z. Berkay Celik}\\
Purdue University
}

\maketitle

\begin{abstract}
Deep learning models have become integral to security applications due to their ability to model complex relationships in data and detect sophisticated threats. However, their complexity makes it difficult to understand how predictions are generated, posing significant challenges for interpretability, particularly in security applications where transparency is critical.
Existing explanation methods, such as visual explanation techniques and post-hoc approaches, suffer from several limitations: reduced faithfulness due to local approximation errors, instability caused by reliance on randomness, and computational inefficiency that hinders real-time usage.
To address these issues, we introduce \system, a self-explainable deep architecture developed for security applications. During training, \system uses a novel mask-based approach to extract informative sub-features from the data and learns prototypes, representative patterns that characterize each class. \system then leverages the prototypes in a dedicated similarity layer at test time to compute similarity scores and generates interpretable explanations without the need for post-hoc analysis.
We evaluate \system across five diverse security scenarios, demonstrating its ability to achieve an average classification accuracy of $97.33\%$ with minimal performance compromise. \system produces deterministic explanations for a fixed trained model and input and substantially reduces explanation latency compared with approximation-based and perturbation-based post-hoc methods.
Through this effort, we extend the applicability of self-explainable AI to security applications, bridging the gap between deep learning performance and the need for explainability in critical scenarios.
\end{abstract}

\section{Introduction}

Deep learning models have achieved strong performance across security tasks such as malware detection, network intrusion detection, and binary code analysis. However, their lack of interpretability limits deployment in critical security settings~\cite{goodman_2017,mink_2023}, where analysts need faithful, understandable, and actionable explanations to support incident response.

To improve interpretability in security applications, prior work has largely adopted post-hoc explainable AI (XAI) methods, where explanations are generated after prediction by analyzing model outputs or internal behavior. These include approximation-based methods (\eg LIME~\cite{lime}, SHAP~\cite{shap}, LEMNA~\cite{lemna}), gradient-based methods (\eg Integrated Gradients~\cite{sundararajan_2017}, Grad-CAM~\cite{selvaraju_2017}, Saliency Maps~\cite{simonyan_2014}), and perturbation-based methods (\eg Occlusion~\cite{matthew_2014}). However, these methods often suffer from low fidelity due to approximation errors, high latency from repeated model queries or gradient computation, white-box access requirements, and instability (where the same input receives different explanations across runs) due to randomness. Recent security-specific methods improve surrogate fidelity~\cite{lemna}, stabilize explanations~\cite{xnids}, or adapt to evolving threats~\cite{yang_2021_cade}, but they remain post-hoc and inherit many of these limitations.

Ante-hoc methods instead embed interpretability directly into the model architecture. Concept-based methods~\cite{cbm,cem,desantis_2024,zabounidis_2023} explain predictions using human-understandable concepts, but require carefully defined and often manually annotated concepts. Prototype-based methods~\cite{kjaersgaard2024pantypes,huang2023evaluation,protopnet,fauvel_lightweight_2023,prototree} explain predictions through representative patterns, but existing methods largely target image data and rely on spatial locality and translational invariance, assumptions that do not hold for many security datasets. Other interpretable architectures, such as transformers~\cite{vaswani_attention_2017} and neural additive models~\cite{agarwal_neural_2021}, provide useful inductive biases, but attention weights are not always faithful explanations of model behavior~\cite{bhusal_sok_2023}.

These limitations motivate a self-explainable architecture for security applications that jointly provides high-fidelity, sparse, stable, and low-latency explanations without relying on post-hoc analysis. We introduce \system, a self-explainable AI framework for security data that performs prediction and explanation jointly at inference time. To address the above requirements, \system learns masks to extract task-relevant sub-features, projects them into an embedding space, and compares them with class-specific prototypes to generate similarity scores. We use these scores both to compute class probabilities and, together with the learned masks, to generate feature importance scores as an explanation.

We evaluate \system on five security applications: PDF malware identification, phishing website detection, network intrusion detection, portable executable (PE) malware classification, and network attack classification. Across these tasks, \system maintains competitive classification performance while producing sparse, stable, and low-latency explanations. Compared with state-of-the-art post-hoc and ante-hoc XAI methods, \system achieves $100\%$ explanation stability and reduces test-time explanation overhead by $3$--$10\times$ on average. Our contributions are:
\begin{itemize}
\item We introduce \system, a self-XAI model for security applications that uses prototype learning to jointly optimize predictive performance and explanation quality.

\item We propose a mask-based sub-feature extraction mechanism that enables \system to learn class-specific prototypes and generate human-understandable feature-importance explanations without post-hoc analysis.

\item We evaluate \system across five security applications and compare it against state-of-the-art XAI methods using established metrics for fidelity, sparsity, stability, and latency. Our results show that \system provides accurate predictions while producing stable, efficient, and high-quality explanations.
\end{itemize}
\section{Related Work}
\label{sec:related_work}

\shortsectionBf{Explanation by Analysis.}
These methods generate explanations through a separate post-hoc analysis step. Approximation-based methods~\cite{lime,shap,lemna} approximate the target model with local explanation models~\cite{minh2022explainable}, but this can reduce faithfulness to the original decision process, add overhead that limits real-time use, and introduce instability from random initialization~\cite{xnids}. In security settings, such limitations can misidentify attack-relevant features, delay response, or produce inconsistent explanations for identical inputs.

Gradient-based methods~\cite{selvaraju_2017,springenberg_2015,smilkov_2017,simonyan_2014,sundararajan_2017} attribute importance using gradients of the output with respect to features. While they can provide detailed attribution signals and are often deterministic, their explanations may be difficult for operators to interpret when features are highly dependent or lack intuitive semantics. They also require white-box access to model parameters, which can be impractical for proprietary or obfuscated security models, and often require domain-specific adaptation to produce human-interpretable feature attributions.

Perturbation-based methods~\cite{fong_2017,matthew_2014,zintgraf_2017} explain predictions by modifying input features or intermediate activations and observing changes in the output. Although conceptually simple, they require repeated model evaluations and can thus incur high computational cost. Depending on the perturbation strategy, they may produce inconsistent explanations due to randomized perturbations.

\shortsectionBf{Explanation by Design.}
In contrast to post-hoc methods, ante-hoc methods build interpretability into the model architecture. SENN~\cite{senn} formalizes self-explaining models around explicitness, faithfulness, and stability, while Neural Additive Models~\cite{agarwal_neural_2021} provide inspectable feature-wise effect functions. Transformers~\cite{vaswani_attention_2017} are often interpreted through attention weights, but attention may not faithfully explain model behavior~\cite{bhusal_sok_2023}. TabNet~\cite{tabnet} uses sequential attention masks for tabular prediction, and InterpreTabNet~\cite{interpretabnet} further improves the interpretability of such masks by encouraging sparsity and diversity; however, these methods are attention-based rather than prototype-based.

Concept-based methods~\cite{cbm,cem,desantis_2024,zabounidis_2023} explain predictions using abstract, human understandable concepts learned from data annotated with labels and concepts. TabCBM~\cite{tabcbm} extends concept-based self-explaining models to tabular data by learning high-level concept explanations with partial or no concept supervision. However, concept-based methods depend on defining a sufficiently complete and interpretable concept set, and supervised variants require time-consuming, task-specific concept annotations~\cite{espinosa2022concept}.

Prototype-based methods explain predictions using representative patterns learned from data. Early prototype-layer models use distances to learned latent prototypes as part of the classifier~\cite{li_case_based_2018}, and ProtoPNet-style architectures \cite{protopnet,prototree,wang_2021,rymarczyk_2021,rymarczyk_2022} learn class-specific prototype parts for image recognition using convolutional feature extractors~\cite{cnn}. LEXNet~\cite{fauvel_lightweight_2023} applies a related explainable-by-design prototype architecture to Internet traffic classification, but still relies on a CNN-based design. More recently, ProtoGate~\cite{protogate} combines prototype-based prediction with global-to-local feature selection, and MEDIC~\cite{medic} learns prototype parts using trainable feature patching and binary or discretized feature subsets. Unlike ProtoGate and MEDIC, which are developed primarily for biomedical tabular data, \system targets feature-vector security tasks and combines prototype-specific masks, class-specific prototype matching, and deterministic feature-importance explanations in a single architecture.
\section{Problem Statement and Requirements}
\label{sec:problem_statement}
We consider a tabular security dataset $\mathcal{D} = \{\mathbf{x}^{(i)}, y^{(i)}\}_{i=1}^{N}$, where $\mathbf{x}^{(i)} \in \mathbb{R}^{d}$ is the $i$th sample and $y^{(i)} \in \{1, \dots, C\}$ is its class label. Each dimension of $\mathbf{x}^{(i)}$ corresponds to a measurable security feature. Our goal is to generate explanations by identifying the features that contribute most to a model prediction, such as whether a sample is malicious or benign. Achieving this goal for tabular security data imposes several design requirements, which we summarize below.

\shortsectionBf{(C$_{1}$) Sub-feature Extraction.} A key challenge is learning prototypes from informative sub-features of tabular security data. Existing prototype-based methods (\eg ProtoPNet~\cite{protopnet}) rely on CNNs to extract patterns. However, security features generally lack spatial locality and translational invariance, making CNN-based pattern extraction unsuitable. We therefore require a mechanism for extracting meaningful feature subsets without assuming image-like structure.

\shortsectionBf{(C$_{2}$) Transparency of the Explanation Method.} Security explanations must be transparent enough for analysts to inspect, trust, and act upon. Complex or non-intuitive explanations can reduce practical utility; for example, gradient-based methods may produce abstract attributions that are difficult for operators to interpret. Thus, the explanation process should be simple, direct, and aligned with the model's prediction mechanism.

\shortsectionBf{(C$_{3}$) Ability to Generate Importance Scores.} Many XAI methods, including LIME, SHAP, and LEMNA, explain predictions using feature-importance scores, while image-based methods often highlight salient regions or prototype-matched patches. For tabular security data, however, visualizing prototypes directly is not meaningful because prototypes are real-valued vectors in an embedding space. The model must therefore convert its internal prototype-based reasoning into human-understandable feature-importance weights.

\shortsectionBf{(C$_{4}$) Multi-Objective Explanation Synthesis.} Effective security explanations must jointly satisfy four objectives: (1) \emph{high fidelity}: explanations reflect the model's true prediction process; (2) \emph{sparsity}: only a small set of features is marked important; (3) \emph{stability}: identical inputs receive consistent explanations across runs; and (4) \emph{low latency}: quick explanations can support real-time use. These requirements are especially important in security, where unstable or slow explanations can delay response or mislead analysts. Unlike image tasks, where explanations may cover broad regions and tolerate small variations, security applications often require concise, consistent, feature-level evidence. We design \system to address these objectives simultaneously.
\section{Methodology}
\label{sec:methodology}
\begin{figure}[t!]
    \centering
    \includegraphics[width=\linewidth]{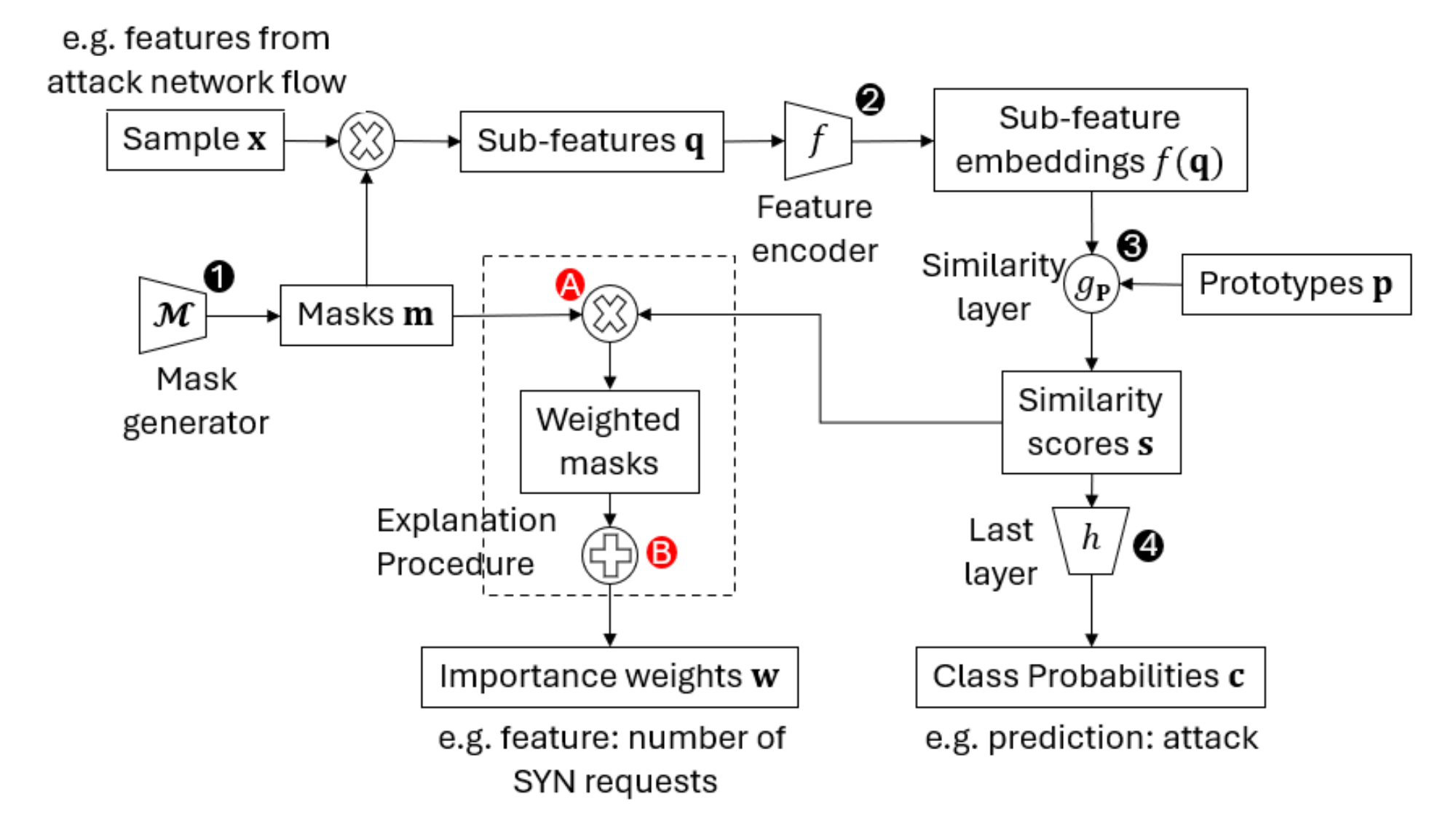}
    \caption{An overview of \system. The input is the sample $\mathbf{x}$, and the outputs are the importance weights $\mathbf{w}$ and the class probabilities $\mathbf{c}$.}
    \label{fig:system}
\end{figure}

We introduce \system, a novel prototype-based self-explainable framework designed for security applications. It is composed of four components as shown in Figure~\ref{fig:system}:
($1$) a mask generator NN $\mathcal{M}$,
($2$) a feature encoder NN $f$,
($3$) a prototype layer $g_\mathbf{P}$, and
($4$) a fully connected layer $h$ that outputs classification scores.
\system also learns a set of prototypes $\mathbf{p}$. Given $C$ classes and $k$ prototypes per class, \system learns $C \times k$ prototypes and generates $C \times k$ masks. The networks $\mathcal{M}$, $f$, and $h$ are not constrained in their architecture. Each can be instantiated with any neural architecture whose input and output dimensions match the required dimensions.

\system uses masks to extract informative sub-features from tabular security data (\textbf{C${1}$}) and a prototype layer to produce feature-importance explanations through a transparent prediction mechanism (\textbf{C${2}$}, \textbf{C${3}$}). A multi-objective loss guides training to satisfy the required explanation properties (\textbf{C${4}$}).
Given an input sample, \system first generates binary masks $\mathbf{m}$ using the mask generator $\mathcal{M}$ (\circled{1}), where each mask has the same dimension as the input. These masks select sub-features, which are then encoded by $f$ into an embedding space (\circled{2}). The prototype layer $g_\mathbf{P}$ computes similarity scores between these embeddings and their corresponding prototypes (\circled{3}). Finally, the fully connected layer $h$ maps the similarity scores to class probabilities using softmax activation~\cite{bridle_1990} (\circled{4}).

In the explanation procedure (\circledRed{\small{A}}, \circledRed{B}), \system leverages the masks and the computed similarity scores $\mathbf{s}$ to generate importance weights $\mathbf{w}$ for each feature such that a higher weight value represents a more important feature.

\subsection{Sub-feature Extraction}
Given a sample $\mathbf{x}^{(i)} \in \mathbb{R}^d$, we extract sub-features and compute their similarity to learned prototypes. A sub-feature is a subset of jointly informative features that indicates a class. For example, in phishing detection, the co-occurrence of links and images may indicate phishing even when either feature alone does not.

\system operates on security datasets where each data point is a vector of real numbers. Unlike images, such data generally lack spatial locality and translational invariance, making CNN-based feature extraction unsuitable. To extract sub-features, \system uses a mask generator $\mathcal{M}$. Given an input vector of length $d$, $\mathcal{M}$ generates a set of masks $\{\mathbf{m}_{j}\}_{j=1}^{C \times k}$. Each mask $\mathbf{m}_{j}$ extracts a sub-feature vector from $\mathbf{x}^{(i)}$ through element-wise multiplication:
\begin{equation}\label{eq:sub-feature}
    \mathbf{q}_{j}^{(i)} = \mathbf{x}^{(i)} \odot \mathbf{m}_{j}
\end{equation}

\shortsectionBf{Binary Masks for Analyst-Interpretable Explanations.} A sigmoid activation restricts the output of $\mathcal{M}$ to $[0, 1]$, but this is insufficient since we require each element to be in $\{ 0, 1 \}$ for intuitive explanations. For example, multiplying the number of links in a PDF by $0.5$ would select half of its value rather than indicate whether the feature is relevant. We therefore encourage each feature to be selected entirely or excluded by introducing the binary-mask loss
\begin{equation}
    \mathcal{L}_{bin} = \sum_{i = 1}^{C \times k} \lVert\mathbf{m}_{i} \odot (1 - \mathbf{m}_{i})\rVert^{2}_{2}
\label{eq:binary_mask_loss}
\end{equation}
Each element-wise product is minimized at $0$ and $1$, thereby encouraging binary masks while preserving differentiability during training.

\shortsectionBf{Learning Sparse Sub-features.} Concise explanations identify only a small subset of the input features and are human-interpretable. Accordingly, each prototype represents an embedding of a compact sub-feature rather than the entire input. We encourage such sparsity in explanations by using the sparsity loss
\begin{equation}
    \mathcal{L}_{spar} = \frac{1}{C \times k} \sum_{i = 1}^{C \times k}\lVert\mathbf{m}_{i}\rVert^{2}_{2}
\label{eq:sparsity_loss}
\end{equation}
For binary masks, $\lVert\mathbf{m}_{i}\rVert_2^2$ equals the number of selected features; minimizing $\mathcal{L}_{spar}$ therefore encourages each mask to select fewer features.

To discourage different prototypes from relying on redundant feature subsets, we introduce the mask-similarity loss
\begin{equation}
    \mathcal{L}_{sim} = \sum_{i = 1}^{C \times k}\sum_{j = i + 1}^{C \times k} \frac{\mathbf{m}_{i} \cdot \mathbf{m}_{j}}{\lVert \mathbf{m}_{i} \rVert \lVert \mathbf{m}_{j} \rVert}
\label{eq:similarity_loss}
\end{equation}
which is the sum of pairwise cosine similarities among the masks. Minimizing it results in masks that are dissimilar.

\subsection{Similarity Score Computation}
\label{sec:similarity}
The feature encoder $f$ maps each sub-feature vector $\mathbf{q}_{j}^{(i)}$ to an embedding. The prototype layer then computes its similarity to the corresponding prototype $\mathbf{p}_{j}$:
\begin{equation}
    \mathbf{s}_{j}^{(i)} = g_{\mathbf{P}_{j}}(f(\mathbf{q}_{j}^{(i)})) = \log\Big(\frac{\lVert f(\mathbf{q}_{j}^{(i)}) - \mathbf{p}_j \rVert_2^2 + 1}{\lVert f(\mathbf{q}_{j}^{(i)}) - \mathbf{p}_j \rVert_2^2 + \varepsilon}\Big)
\label{eq:similarity}
\end{equation}
where $0 < \varepsilon < 1$ prevents division by zero. This score decreases with the distance between the sub-feature embedding and its prototype; therefore, a higher score indicates a closer match.

To learn a useful embedding space, we minimize the cluster loss
\begin{equation}
\mathcal{L}_{cls} = \frac{1}{N} \sum_{i = 1}^{N} \min_{j: \mathbf{p}_{j} \in \mathbf{P}_{y^{(i)}}} \lVert f(\mathbf{q}_{j}^{(i)}) - \mathbf{p}_{j} \rVert_{2}^{2}
\label{eq:cluster_loss}
\end{equation}
where $\mathbf{P}_{y^{(i)}}$ denotes the prototypes associated with the true class of $\mathbf{x}^{(i)}$. This loss encourages at least one class-associated sub-feature embedding to be close to its corresponding prototype. We enforce Lipschitz continuity of $f$ following~\cite{lipschitz}, limiting changes in the learned embeddings (and consequently the explanations) under small input perturbations.

\begin{figure*}[t!]
\centering
\includegraphics[width=\textwidth]{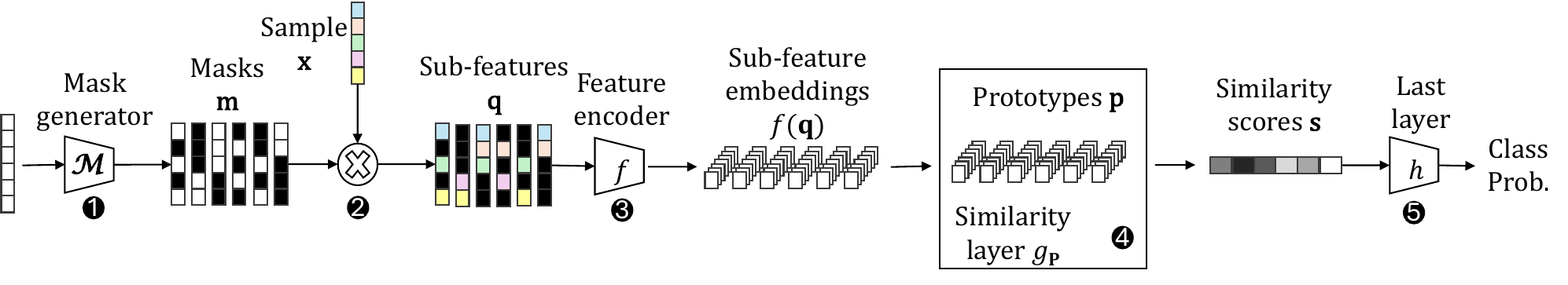}
\caption{Illustration of \system classifying a five-dimensional sample using a seven-dimensional embedding and three prototypes per class ($k = 3$). Colors identify input features; embeddings and prototypes are uncolored because they reside in the learned embedding space. Lighter shading in the similarity scores indicates greater prototype-embedding similarity.}
\label{fig:example}
\end{figure*}

\subsection{Last Layer Classification}
The final layer $h$ maps similarities $\mathbf{s}^{(i)}$ to class probabilities:
\begin{equation}
\label{eq:last_layer}
\mathbf{c}^{(i)} = \text{Softmax}(h(\mathbf{s}^{(i)}))
\end{equation}
Because $h$ receives only prototype-similarity scores, predictions are determined entirely by prototype matches, directly linking classification and explanation. The predicted class is
$o^{(i)}=\arg\max_j c_j^{(i)}$.

Figure~\ref{fig:example} illustrates \system's classification process with three prototypes per class ($k=3$). The prototypes and masks are learned during training and fixed thereafter. Thus, all test samples use the same prototype-specific masks and prototypes; neither is generated per sample.

\subsection{Training Procedure}\label{sec:training}
We divide \system's training process into two phases:

\shortsectionBf{Phase 1: Joint Learning.}
In this phase, the mask generator, feature encoder, and prototypes are jointly trained to learn informative masks, prototypes, and embeddings while maintaining predictive performance. We use cross-entropy loss
\begin{equation}
\mathcal{L}_{xe} = -\frac{1}{N}\sum_{i = 1}^{N} \sum_{j = 1}^{C} \log(c_{j}^{(i)}) \cdot \mathbbm{1}(y^{(i)} = j)
\label{eq:crossentropy_loss}
\end{equation}
where $c_{j}^{(i)}$ is the predicted probability of class $j$ for $\mathbf{x}^{(i)}$.

During this phase, the last-layer weights are fixed to guide prototype learning. Let $\mathbf{W}_{h}$ denote the weight matrix of $h$, where $W_{h}^{jl}$ connects the similarity score of prototype $\mathbf{p}_{j}$ to the logit for class $l$. The weight is positive when $\mathbf{p}_j$ belongs to class $l$ and negative otherwise. Thus, matching a prototype increases its associated class logit while decreasing the logits of other classes. We set
\begin{equation}
\label{eq:weight_initialization}
\textbf{W}_{h}^{jl} = \begin{cases}
1 &\text{if } \mathbf{p}_{j} \in \mathbf{P}_{l}\\
-0.5 &\text{if } \mathbf{p}_{j} \not\in \mathbf{P}_{l}
\end{cases}
\end{equation}
The negative weights ensure that when a sample matches a prototype from one class, its scores for competing classes decrease. This helps the model distinguish between classes more clearly.

We train the mask generator, feature encoder, and prototypes jointly while keeping the last layer $h$ fixed. The joint objective is
\begin{equation}
    \mathcal{L}_{joint} =  \lambda_{0} \mathcal{L}_{xe} + \lambda_{1} \mathcal{L}_{spar} + \lambda_{2} \mathcal{L}_{cls} + \lambda_{3} \mathcal{L}_{bin} + \lambda_{4} \mathcal{L}_{sim}
\label{eq:loss}
\end{equation}
where the coefficients $\lambda$ are tunable hyperparameters.

\shortsectionBf{Phase 2: Last Layer Fine-tuning.}
After joint training, we freeze the mask generator, feature encoder, and prototypes and optimize only the last layer $h$. To improve classification while suppressing connections between prototypes and unrelated classes, we use
\begin{equation}
\mathcal{L}_{reg} = \sum_{l = 1}^{C} \sum_{\mathbf{p}_j \notin \mathbf{P}_{l}} \lvert \textbf{W}_{h}^{jl} \rvert
\label{eq:reg}
\end{equation}
The fine-tuning objective is
\begin{equation}
\mathcal{L}_{last} = \mathcal{L}_{xe} + \lambda \mathcal{L}_{reg}
\end{equation}
where $\lambda$ controls the regularization strength. By shrinking off-class weights toward zero, $\mathcal{L}_{reg}$ encourages predictions to rely on similarities to prototypes of the predicted class rather than on the absence of similarities to other-class prototypes. Thus, the model primarily relies on positive evidence in its predictions.

\begin{figure}[t!]
\centering
\includegraphics[width=\linewidth]{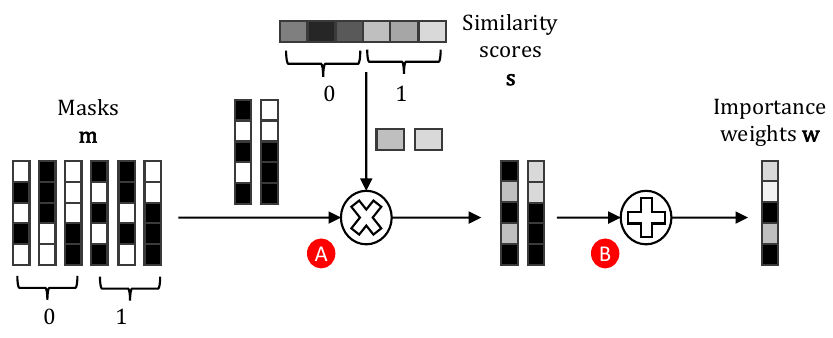}
\caption{The explanation process of \system. Here, the number of prototypes per class $k = 3$ and the number of highest similarity scores to consider $n = 2$.}
\label{fig:explain}
\end{figure}

\subsection{Explanation Procedure}
\label{sec:explain}

\system generates a feature-importance vector $\mathbf{w}$, where larger values indicate greater importance. Figure~\ref{fig:explain} shows the explanation process in detail for $k = 3$ and $n = 2$. Lighter shades represent higher similarity scores. The figure assumes that the sample has been classified as class $1$, and therefore it considers the highest $2$ similarity scores for class $1$ and their corresponding masks to generate importance weights. This is consistent with our requirement that higher similarity scores translate to higher feature importance weights.

Algorithm~\ref{alg:explain} details this procedure. Given $n\in\{1,\ldots,k\}$, \system extracts the sub-features (line~1), computes their prototype-similarity scores (line~2), and obtains the predicted class (lines~3--4). It then selects the $n$ highest-scoring prototypes associated with that class (line~5) and computes $\mathbf{w}$ as a similarity-weighted sum of their masks (line~6, \circledRed{\small{A}}, \circledRed{B}).

\begin{algorithm}[t!]
\small
\caption{Explanation procedure for \system.}
\label{alg:explain}
\begin{algorithmic}[1]
\Require Sample $\mathbf{x}$, $n \in \mathbb{Z} : n \in [1, k]$
\Ensure Predicted class $o$, importance scores $\mathbf{w}$
\State $\mathbf{q} \gets \mathbf{x} \odot \mathbf{m}$
\State $\mathbf{s} = g_{\mathbf{P}}(f(\mathbf{q})) = \log\Big(\frac{\lVert f(\mathbf{q}) - \mathbf{p} \rVert_2^2 + 1}{\lVert f(\mathbf{q}) - \mathbf{p} \rVert_2^2 + \varepsilon}\Big)$
\State $\mathbf{c} = \text{Softmax}(h(\mathbf{s}))$
\State $o = \argmax_{j} \mathbf{c}_{j}$
\State Let $I$ be the indices of top $n$ similarity scores corresponding to class $o$.
\State $\mathbf{w} = \sum_{i \in I}{s_{i} \cdot \mathbf{m}_{i}}$
\State \Return $o$, $\mathbf{w}$
\end{algorithmic}
\end{algorithm}

\shortsectionBf{Deterministic Explanations.}
For a fixed model and input, Algorithm~\ref{alg:explain} is deterministic and therefore always produces the same explanation. This avoids the run-to-run variability caused by stochastic post-hoc explanation procedures.

\shortsectionBf{Real-Time Explanations.}
Prediction and explanation share a single forward pass. Explanations require selecting the highest-scoring prototypes and aggregating their masks, without training surrogate models or perturbing inputs.
\section{Experiments}\label{sec:evaluation}

We evaluate \system on five security datasets and show that it maintains competitive classification performance, and that its explanations are high-fidelity, sparse, stable, and low-latency.

\subsection{Experimental Setup}
\shortsectionBf{Datasets.}
We evaluate \system on five security datasets: PDF malware identification~\cite{pdfmalware,mimicus}, phishing website detection~\cite{phishing}, network intrusion detection~\cite{sarhan_2022}, PE malware classification~\cite{bodmas}, and network attack classification~\cite{tavallaee_2009_detailed}. These datasets cover binary and multi-class tasks with $2$--$5$ classes, $10$K--$20$K samples, and $38$--$135$ features, and have been used in prior security-XAI studies~\cite{lemna,sheatsley_2021,xnids}. The PDF malware dataset contains $135$ structural PDF features from benign and malicious files; the phishing dataset contains $38$ URL/webpage features from phishing and legitimate websites; the network intrusion dataset contains $39$ NetFlow features sampled from benign and attack traffic; the PE malware dataset uses BODMAS malware-family features, retaining families with at least $3{,}000$ samples and the top $100$ ANOVA-ranked non-constant features; and the network attack dataset uses NSL-KDD flow features for benign traffic and attack classes with at least $3{,}000$ samples. For all datasets, we use an $80/20$ stratified train-test split and standardize features to zero mean and unit variance. We open-source preprocessing scripts and dataset subsets. Please see Appendix~\ref{app:dataset_details} for details of the datasets.

\shortsectionBf{Baselines.}
We compare \system with representative post-hoc and ante-hoc XAI methods. The post-hoc baselines include ($1$) approximation-based methods, LIME \cite{lime} and SHAP~\cite{shap}; the security-specific improvement LEMNA~\cite{lemna}; ($2$) gradient-based methods, Integrated Gradients (IG)~\cite{sundararajan_2017}, Guided Grad-CAM (GGC)~\cite{selvaraju_2017}, and Saliency Maps (SM)~\cite{simonyan_2014}; and ($3$) a perturbation-based method, Occlusion (Occl.)~\cite{matthew_2014}. The ante-hoc baselines include ProtoPNet (PPN)~\cite{protopnet} and a Transformer architecture (Tran.)~\cite{vaswani_attention_2017}. Since ProtoPNet is designed for images, we adapt it using 1D convolutions and derive feature importance using gradients. For the Transformer, we use averaged attention weights as feature-importances. We additionally compare with xNIDS~\cite{xnids} on network intrusion detection.

\shortsectionBf{Evaluation Metrics.}
\label{sec:evaluation_metrics}
We measure classification performance using accuracy, precision, recall, and false positive rate (FPR), and explanation quality using fidelity, sparsity, stability, and latency metrics.

\shortsectionEmph{Fidelity.}
Fidelity measures whether an explanation identifies features that actually drive the model's prediction. Following prior work~\cite{lemna,xnids,wojciech_lerfmorf_2017}, we use three feature-perturbation tests. For a sample $\mathbf{x}$, let $\mathbf{F}_{\mathbf{x}}$ denote the top features selected by an explanation method, with $\lvert \mathbf{F}_{\mathbf{x}} \rvert$ being the explanation size.

\begin{enumerate}[topsep=0pt, itemsep=0.5pt]
\item \textit{Synthetic Test.} We preserve the selected features $\mathbf{F}_{\mathbf{x}}$ in $\mathbf{x}$ and randomize all remaining features. A faithful explanation should preserve the original prediction under this perturbation. This test is analogous to LeRF~\cite{wojciech_lerfmorf_2017}.

\item \textit{Feature Augmentation Test.} We insert the selected features $\mathbf{F}_{\mathbf{x}}$ from $\mathbf{x}$ into a sample from a different class to obtain $\mathbf{x}_{s}$. A faithful explanation should cause $\mathbf{x}_{s}$ to be classified as $\mathbf{x}$'s original class.

\item \textit{Feature Deduction Test.} We randomize the selected features $\mathbf{F}_{\mathbf{x}}$ in $\mathbf{x}$ while preserving all other features. A faithful explanation should change the original prediction. This test is analogous to MoRF~\cite{wojciech_lerfmorf_2017}.
\end{enumerate}

\shortsectionEmph{Sparsity.}
Sparse explanations are easier to inspect because they identify only a small set of important features~\cite{warnecke_2020}. We measure sparsity using the Mean Around Zero (MAZ) curve: after normalizing absolute feature-importance scores to $[0,1]$, MAZ measures how much score mass lies near zero. A steeper curve near zero indicates a sparser explanation.

\shortsectionEmph{Stability.}
Stability measures whether an explanation method identifies the same important features across runs~\cite{warnecke_2020}. For two runs with top-$N$ feature sets $T_{i}$ and $T_{j}$, we compute $|T_i \cap T_j|/N$; values closer to $1.0$ indicate more stable explanations.

\shortsectionBf{Implementation.}
We implement \system in PyTorch~\cite{pytorch} with the Adam optimizer~\cite{adam}. We use $k=10$ prototypes per class for all datasets. The feature encoder $f$ is an MLP for the PDF malware, phishing website, network intrusion, and PE malware datasets, and an LSTM for the network attack dataset to model temporal relationships. We select loss coefficients, learning rates, batch sizes, embedding dimensions, and network widths using K-fold cross-validation. Experiments are run on a laptop with an AMD Ryzen 7 CPU, an NVIDIA GeForce RTX 3050Ti GPU, and 16GB RAM. For the baselines, we use published implementations when available~\cite{lime_code,shap_code,lemna_code,captum,xnids_code}, and include preprocessing scripts and baseline configurations in our released code\footnote{\url{https://github.com/ashreeku/XSec}}. \system hyperparameters are reported in Appendix~\ref{app:hyperparameters}.

\begin{table}[t!]
\caption{Classification performance across datasets.}
\vspace{1em}
\label{tab:performance}
\centering
\setlength{\tabcolsep}{3pt}
\resizebox{\columnwidth}{!}{
\begin{threeparttable}
\begin{tabular}{|c|c|ccccc|}
\hline
\multirow{2}{*}{\textbf{Classifier}} & \multirow{2}{*}{\textbf{Metric$^\dagger$}} & \multicolumn{5}{c|}{\textbf{Dataset}}                                                                                                           \\ \cline{3-7} 
                                     &                                  & \multicolumn{1}{c|}{\textbf{\begin{tabular}[c]{@{}c@{}}PDF\\ Malware\end{tabular}}} & \multicolumn{1}{c|}{\textbf{\begin{tabular}[c]{@{}c@{}}Website\\ Phishing\end{tabular}}} & \multicolumn{1}{c|}{\textbf{\begin{tabular}[c]{@{}c@{}}Network\\ Intrusion\end{tabular}}} & \multicolumn{1}{c|}{\textbf{\begin{tabular}[c]{@{}c@{}}PE\\ Malware\end{tabular}}} & \multicolumn{1}{c|}{\textbf{\begin{tabular}[c]{@{}c@{}}Network\\ Attack\end{tabular}}} \\ \hline\hline
\multirow{4}{*}{\textbf{\system}}    & \textbf{Acc.} & \multicolumn{1}{c|}{98.95\%} & \multicolumn{1}{c|}{93.85\%} & \multicolumn{1}{c|}{99.15\%} & \multicolumn{1}{c|}{94.79\%} & 99.92\% \\
                                     & \textbf{Pre.} & \multicolumn{1}{c|}{99.30\%} & \multicolumn{1}{c|}{93.03\%} & \multicolumn{1}{c|}{98.91\%} & \multicolumn{1}{c|}{95.30\%} & 99.92\% \\
                                     & \textbf{Rec.} & \multicolumn{1}{c|}{98.60\%} & \multicolumn{1}{c|}{94.80\%} & \multicolumn{1}{c|}{99.40\%} & \multicolumn{1}{c|}{95.38\%} & 99.92\% \\
                                     & \textbf{FPR}  & \multicolumn{1}{c|}{0.70\%}  & \multicolumn{1}{c|}{7.10\%}  & \multicolumn{1}{c|}{1.10\%}  & \multicolumn{1}{c|}{1.32\%}  & 0.03\%  \\ \hline
\multirow{4}{*}{\textbf{MLP / LSTM}} & \textbf{Acc.} & \multicolumn{1}{c|}{99.85\%} & \multicolumn{1}{c|}{96.10\%} & \multicolumn{1}{c|}{99.42\%} & \multicolumn{1}{c|}{95.72\%} & 99.92\% \\
                                     & \textbf{Pre.} & \multicolumn{1}{c|}{99.90\%} & \multicolumn{1}{c|}{96.19\%} & \multicolumn{1}{c|}{99.06\%} & \multicolumn{1}{c|}{96.13\%} & 99.92\% \\
                                     & \textbf{Rec.} & \multicolumn{1}{c|}{99.80\%} & \multicolumn{1}{c|}{96.00\%} & \multicolumn{1}{c|}{99.80\%} & \multicolumn{1}{c|}{96.21\%} & 99.92\% \\
                                     & \textbf{FPR}  & \multicolumn{1}{c|}{0.10\%}  & \multicolumn{1}{c|}{3.80\%}  & \multicolumn{1}{c|}{0.95\%}  & \multicolumn{1}{c|}{1.09\%}  & 0.03\%  \\ \hline
\multirow{4}{*}{\textbf{CNN}}        & \textbf{Acc.} & \multicolumn{1}{c|}{99.85\%} & \multicolumn{1}{c|}{96.45\%} & \multicolumn{1}{c|}{99.45\%} & \multicolumn{1}{c|}{96.53\%} & 99.96\% \\
                                     & \textbf{Pre.} & \multicolumn{1}{c|}{100.0\%} & \multicolumn{1}{c|}{96.87\%} & \multicolumn{1}{c|}{99.06\%} & \multicolumn{1}{c|}{96.99\%} & 99.96\% \\
                                     & \textbf{Rec.} & \multicolumn{1}{c|}{99.70\%} & \multicolumn{1}{c|}{96.00\%} & \multicolumn{1}{c|}{99.85\%} & \multicolumn{1}{c|}{96.93\%} & 99.96\% \\
                                     & \textbf{FPR}  & \multicolumn{1}{c|}{0.00\%}  & \multicolumn{1}{c|}{3.10\%}  & \multicolumn{1}{c|}{0.95\%}  & \multicolumn{1}{c|}{0.89\%}  & 0.01\%  \\ \hline
\multirow{4}{*}{\textbf{ProtoPNet}}  & \textbf{Acc.} & \multicolumn{1}{c|}{99.80\%} & \multicolumn{1}{c|}{96.30\%} & \multicolumn{1}{c|}{99.47\%} & \multicolumn{1}{c|}{96.56\%} & 100.0\% \\
                                     & \textbf{Pre.} & \multicolumn{1}{c|}{100.0\%} & \multicolumn{1}{c|}{96.30\%} & \multicolumn{1}{c|}{99.16\%} & \multicolumn{1}{c|}{96.89\%} & 100.0\% \\
                                     & \textbf{Rec.} & \multicolumn{1}{c|}{99.60\%} & \multicolumn{1}{c|}{96.30\%} & \multicolumn{1}{c|}{99.80\%} & \multicolumn{1}{c|}{96.94\%} & 100.0\% \\
                                     & \textbf{FPR}  & \multicolumn{1}{c|}{0.00\%}  & \multicolumn{1}{c|}{3.70\%}  & \multicolumn{1}{c|}{0.85\%}  & \multicolumn{1}{c|}{0.88\%}  & 0.00\%  \\ \hline
\multirow{4}{*}{\textbf{Transformer}}& \textbf{Acc.} & \multicolumn{1}{c|}{97.55\%} & \multicolumn{1}{c|}{94.40\%} & \multicolumn{1}{c|}{96.85\%} & \multicolumn{1}{c|}{95.70\%} & 98.00\% \\
                                     & \textbf{Pre.} & \multicolumn{1}{c|}{98.57\%} & \multicolumn{1}{c|}{92.20\%} & \multicolumn{1}{c|}{94.58\%} & \multicolumn{1}{c|}{95.94\%} & 98.00\% \\
                                     & \textbf{Rec.} & \multicolumn{1}{c|}{96.50\%} & \multicolumn{1}{c|}{97.00\%} & \multicolumn{1}{c|}{99.40\%} & \multicolumn{1}{c|}{96.13\%} & 98.0\% \\
                                     & \textbf{FPR}  & \multicolumn{1}{c|}{1.40\%}  & \multicolumn{1}{c|}{8.20\%}  & \multicolumn{1}{c|}{5.70\%}  & \multicolumn{1}{c|}{1.10\%}  & 0.67\%  \\ \hline
\end{tabular}
\begin{tablenotes}[flushleft]
\item $^\dagger$ Acc.: accuracy, Pre.: precision, Rec.: recall, and FPR: false positive rate.
\item For precision, recall, and FPR, mean value is reported for multi-class datasets.
\end{tablenotes}
\end{threeparttable}}
\end{table}

\begin{figure*}[t!]
\centering
\includegraphics[width=\textwidth]{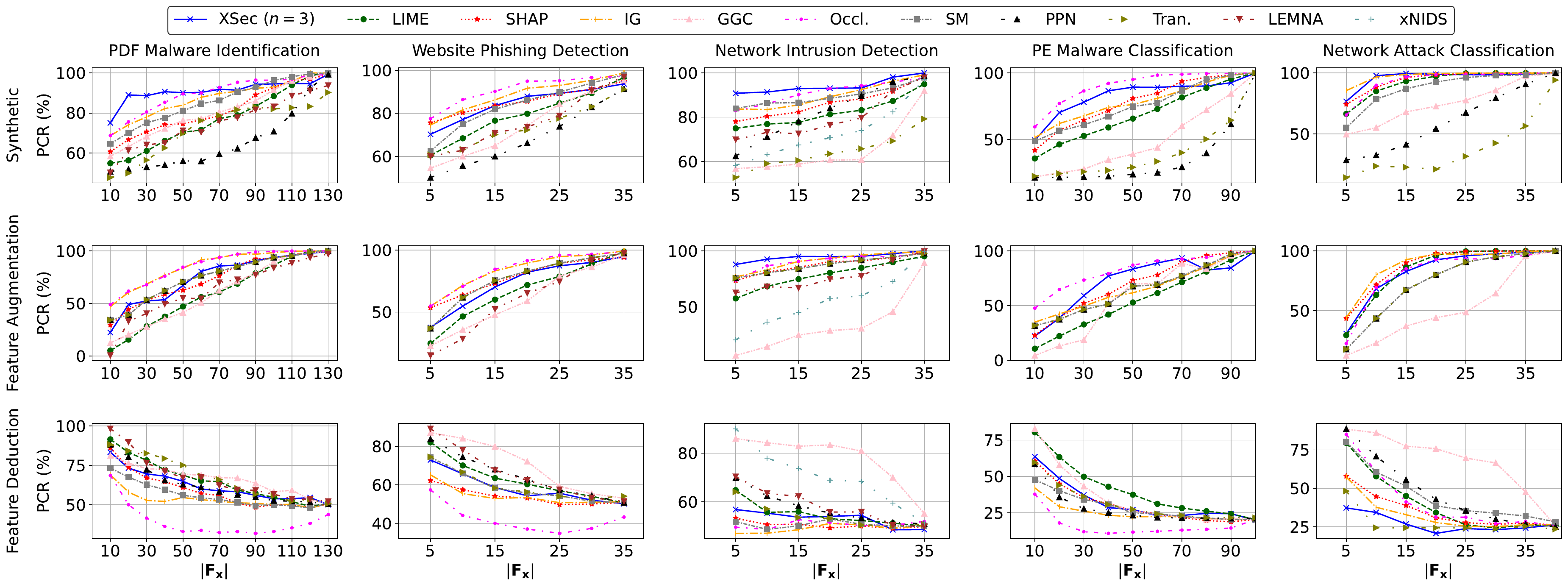}
\caption{Comparison of \system with baseline explanation methods under three fidelity tests across all datasets. Higher PCR is better for the synthetic and feature augmentation tests; lower PCR is better for the feature deduction test.}
\label{fig:fidelity_tests}
\end{figure*}

\subsection{Classification Performance}
\label{sec:performance}
We first evaluate whether \system preserves classification performance despite its architectural constraints. For each dataset, we compare \system against tuned neural baselines: an MLP for PDF malware, phishing website, network intrusion, and PE malware classification; an LSTM for network attack classification; and CNN, Transformer, and ProtoPNet models across all datasets.

Table~\ref{tab:performance} shows that \system achieves competitive performance across all tasks. Compared with the strongest neural baselines, \system incurs only a small accuracy loss on PDF malware, phishing website, and PE malware classification, while achieving near-identical performance on network intrusion and network attack classification. These results show that \system provides self-explainable predictions with minimal compromise in predictive performance.

\subsection{Explainability Results}\label{sec:fidelity}

\shortsectionBf{Fidelity.}
For each test sample, we apply the three fidelity tests and report the positive classification rate (PCR) as a function of $\lvert \mathbf{F}_{\mathbf{x}} \rvert$. PCR is the fraction of modified samples classified as the original prediction. Higher PCR indicates better fidelity for the synthetic and feature augmentation tests, while lower PCR indicates better fidelity for the feature deduction test.

Figure~\ref{fig:fidelity_tests} shows that \system provides consistently strong fidelity across datasets. In the synthetic test, \system performs best on PDF malware, network intrusion, and network attack classification, and remains comparable on phishing and PE malware. In the feature augmentation test, \system is strongest on network intrusion and competitive elsewhere. In the feature deduction test, \system remains comparable to other methods, although Occlusion performs best. \system consistently outperforms LIME, SHAP, Saliency Maps, ProtoPNet, and Transformer.

\begin{figure*}[t!]
\centering
\includegraphics[width=\textwidth]{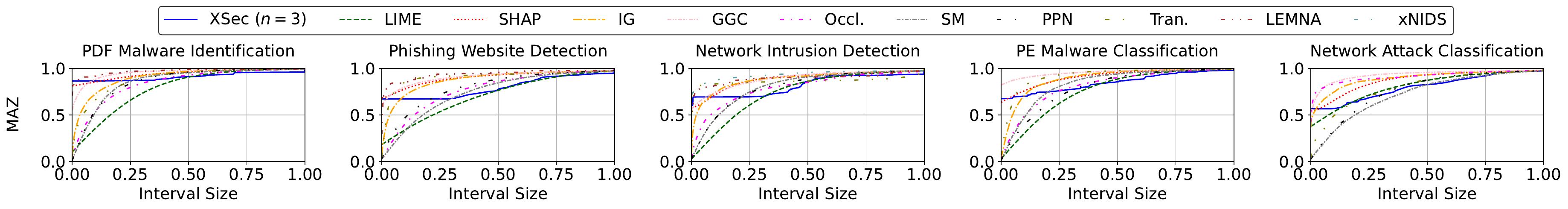}
\caption{The Mean Around Zero (MAZ) sparsity curves. A steeper curve closer to $0$ indicates a sparser explanation method.}
\label{fig:sparsity}
\end{figure*}

\shortsectionBf{Sparsity.}
Figure~\ref{fig:sparsity} shows the MAZ curves for each explanation method, and Table~\ref{tab:sparsity_auc} reports their AUC values. Curves that rise sharply near zero indicate sparser explanations. \system consistently produces steep MAZ curves and high AUC scores across datasets, indicating that its explanations place importance on a small set of features.

Some baselines achieve higher sparsity than \system on individual datasets, but sparsity must be interpreted alongside fidelity: an explanation that is sparse yet does not reflect the model's decision is not useful. For example, LEMNA, SHAP, GGC, Transformer, and xNIDS sometimes produce sparser explanations, but they generally show lower fidelity in Figure~\ref{fig:fidelity_tests}. Conversely, Occlusion is less sparse but performs strongly in the feature deduction test. Overall, \system provides a favorable balance between sparsity and fidelity across datasets.

\begin{table}[t!]
\centering
\caption{Area Under the Curve of the MAZ sparsity curves. A score closer to $1.0$ along with a steep rise near $0$ in Figure~\ref{fig:sparsity} indicates sparser explanations.}
\vspace{1em}
\label{tab:sparsity_auc}
\setlength{\tabcolsep}{3pt}
\resizebox{\columnwidth}{!}{
\begin{threeparttable}
\begin{tabular}{|c|c|c|c|c|c|c|c|c|c|c|}
\hline
 & \textbf{\system} & \textbf{LIME} & \textbf{SHAP} & \textbf{LEMNA} & \textbf{IG} & \textbf{GGC} & \textbf{Occl.} & \textbf{SM} & \textbf{PPN} & \textbf{Tran.} \\ \hline\hline
\textbf{\makecell{PDF\\ Malware}}       & $0.909$           & $0.785$         & $0.942$         & $0.970$   &   $0.904$   &    $0.952$   &   $0.841$ & $0.859$ & $0.858$ & $0.884$    \\ \hline
\textbf{\makecell{Phishing\\Website}}  & $0.793$           & $0.704$         & $0.901$         & $0.935$   &   $0.877$   &    $0.909$   &   $0.770$ & $0.743$ & $0.795$ & $0.904$    \\ \hline
\textbf{\makecell{Network\\ Intrusion}} & $0.819$           & $0.741$         & $0.886$         & $0.893$   &   $0.883$   &    $0.890$   &   $0.796$ & $0.805$  & $0.799$  & $0.863$   \\ \hline
\textbf{\makecell{PE\\ Malware}}        & $0.849$           & $0.761$         & $0.912$         & $-^\dagger$       &   $0.892$   &    $0.956$   &   $0.813$ & $0.838$  & $0.795$  & $0.921$   \\ \hline
\textbf{\makecell{Network\\ Attack}}    & $0.798$           & $0.809$         & $0.874$         & $-^\dagger$       &   $0.891$   &    $0.928$   &   $0.911$ & $ 0.726$ & $ 0.760$ & $ 0.793$    \\ \hline
\end{tabular}
\begin{tablenotes}[flushleft]
\item $^\dagger$ LEMNA does not support multi-class datasets.
\end{tablenotes}
\end{threeparttable}}
\end{table}

\shortsectionBf{Stability.}
We evaluate stability by running each explanation method with two different seeds and measuring the average overlap between the top-$10$ features selected in each run. The results are reported in Table~\ref{tab:stability}. \system is deterministic for a fixed trained model and input, and therefore achieves a stability score of $1.0$. The deterministic baselines---IG, GGC, Saliency Maps, Occlusion, ProtoPNet, and Transformer---also achieve perfect stability. In contrast, post-hoc methods such as LIME, SHAP, LEMNA, and xNIDS produce lower stability because their explanations depend on stochastic perturbation or sampling procedures.

\begin{table}[t!]
\centering
\caption{Stability of XAI methods. A score closer to $1.0$ indicates higher stability.}
    \vspace{1em}
\label{tab:stability}
\centering
\setlength{\tabcolsep}{3pt}
\resizebox{\columnwidth}{!}{
\begin{threeparttable}
\begin{tabular}{|c|c|c|c|c|c|c|c|c|c|c|}
\hline
                      & \textbf{\system} & \textbf{LIME} & \textbf{SHAP} & \textbf{LEMNA} & \textbf{IG} & \textbf{GGC} & \textbf{Occl.} & \textbf{SM} & \textbf{PPN} & \textbf{Tran.} \\ \hline \hline
\textbf{\makecell{PDF\\Malware}}       & $1.0$           & $0.116$       & $0.170$        & $0.381$   &    $1.0$    &     $1.0$    &    $1.0$   &   $1.0$    &    $1.0$   &   $1.0$ \\ \hline
\textbf{\makecell{Phishing\\ Website}}  & $1.0$           & $0.559$       & $0.771$        & $0.704$   &    $1.0$    &     $1.0$    &    $1.0$  &   $1.0$    &    $1.0$   &   $1.0$ \\ \hline
\textbf{\makecell{Network\\ Intrusion}} & $1.0$           & $0.419$       & $0.532$        & $0.740$   &    $1.0$    &     $1.0$    &    $1.0$  &   $1.0$    &    $1.0$   &   $1.0$ \\ \hline
\textbf{\makecell{PE\\ Malware}}        & $1.0$           & $0.192$       & $0.229$        & $-$   &    $1.0$    &     $1.0$    &    $1.0$  &   $1.0$    &    $1.0$   &   $1.0$ \\ \hline
\textbf{\makecell{Network\\ Attack}}    & $1.0$           & $0.560$       & $0.586$        & $-$       &    $1.0$    &     $1.0$    &    $1.0$  &   $1.0$    &    $1.0$   &   $1.0$ \\ \hline
\end{tabular}
\end{threeparttable}}
\end{table}

\shortsectionBf{Latency.}
We measure test-set explanation time by running each method on test samples individually, without batching or parallelization, under the same experimental setup. Each experiment is repeated three times, and Table~\ref{tab:timing} reports the mean and standard deviation. \system has low test-time latency because prediction and explanation share a single forward pass. In contrast, approximation-based methods such as LIME, SHAP, and LEMNA require repeated perturbation and model evaluation, resulting in substantially higher latency. xNIDS is the slowest method on network intrusion detection, requiring $7{,}200 \pm 600$ seconds.

\begin{table}[t!]
\caption{Mean and standard deviation of latency in seconds over three runs.}
\vspace{1em}
\label{tab:timing}
\centering
\resizebox{\columnwidth}{!}{
\begin{tabular}{|c|c|c|c|c|c|c|c|c|c|c|}
\hline
 & \textbf{\system}                                         & \textbf{LIME}                                              & \textbf{SHAP}                                               & \textbf{LEMNA}     &   \textbf{IG} &   \textbf{GGC} & \textbf{Occl.} & \textbf{SM} & \textbf{PPN} & \textbf{Tran.}  \\ \hline\hline
\textbf{\makecell{PDF\\ Malware}}       & \makecell{$3.37$\\ $\pm 0.18$} & \makecell{$193.85$\\ $\pm 11.11$} & \makecell{$401.68$\\ $\pm 14.02$}  & \makecell{$2296.84$\\ $\pm 9.37$}  & \makecell{$10.86$\\ $\pm 0.29$}  & \makecell{$2.60$\\ $\pm 0.09$}  & \makecell{$74.73$\\ $\pm 0.16$} & \makecell{$0.74$ \\ $\pm 0.01$} & \makecell{$18.75$ \\ $\pm 1.91$} & \makecell{$7.52$ \\ $\pm 0.03$} \\ \hline
\textbf{\makecell{Phishing\\ Website}}  & \makecell{$1.38$\\ $\pm 0.14$} & \makecell{$18.19$\\ $\pm 0.30$}   & \makecell{$42.82$\\ $\pm 0.53$}    & \makecell{$516.77$\\ $\pm 8.65$}    & \makecell{$10.60$\\ $\pm 0.10$}    & \makecell{$2.17$\\ $\pm 0.02$}    & \makecell{$20.24$\\ $\pm 0.34$}    & \makecell{$0.70$\\ $\pm 0.01$}    & \makecell{$3.29$\\ $\pm 0.01$}    & \makecell{$6.43$\\ $\pm 0.05$} \\ \hline
\textbf{\makecell{Network\\ Intrusion}} & \makecell{$4.59$\\ $\pm 0.27$} & \makecell{$40.35$\\ $\pm 4.64$}   & \makecell{$293.69$\\ $\pm 37.21$}  & \makecell{$1060.75$\\ $\pm 44.99$}  & \makecell{$27.63$\\ $\pm 0.57$}  & \makecell{$5.46$\\ $\pm 0.07$}  & \makecell{$42.17$\\ $\pm 0.27$}  & \makecell{$1.49$\\ $\pm 0.05$}  & \makecell{$6.39$\\ $\pm 0.01$}  & \makecell{$12.87$\\ $\pm 0.76$} \\ \hline
\textbf{\makecell{PE\\ Malware}}        & \makecell{$6.60$\\ $\pm 0.77$} & \makecell{$357.03$\\ $\pm 18.02$} & \makecell{$1871.55$\\ $\pm 34.23$} & $-$ & \makecell{$21.15$\\ $\pm 0.33$} & \makecell{$4.76$\\ $\pm 0.03$} & \makecell{$111.51$\\ $\pm 0.12$} & \makecell{$1.54$\\ $\pm 0.06$} & \makecell{$150.27$\\ $\pm 3.00$} & \makecell{$14.25$\\ $\pm 0.57$} \\ \hline
\textbf{\makecell{Network\\ Attack}}        & \makecell{$5.22$\\ $\pm 0.35$} & \makecell{$28.11$\\ $\pm 0.24$} & \makecell{$171.16$\\ $\pm 2.59$} & $-$ & \makecell{$13.45$\\ $\pm 0.20$} & \makecell{$2.66$\\ $\pm 0.10$} & \makecell{$54.37$\\ $\pm 0.46$} & \makecell{$2.40$\\ $\pm 0.01$} & \makecell{$1.97$\\ $\pm 0.07$} & \makecell{$6.76$\\ $\pm 0.15$} \\ \hline
\end{tabular}
}
\end{table}

\subsection{Hyperparameter Sensitivity}\label{sec:hyperparameter_sensitivity}

We perform experiments for hyperparameter sensitivity of \system with respect to the number of prototypes per class ($k$) and the number of similarity scores ($n$).

\begin{figure*}[t!]
\centering
\includegraphics[width=\textwidth]{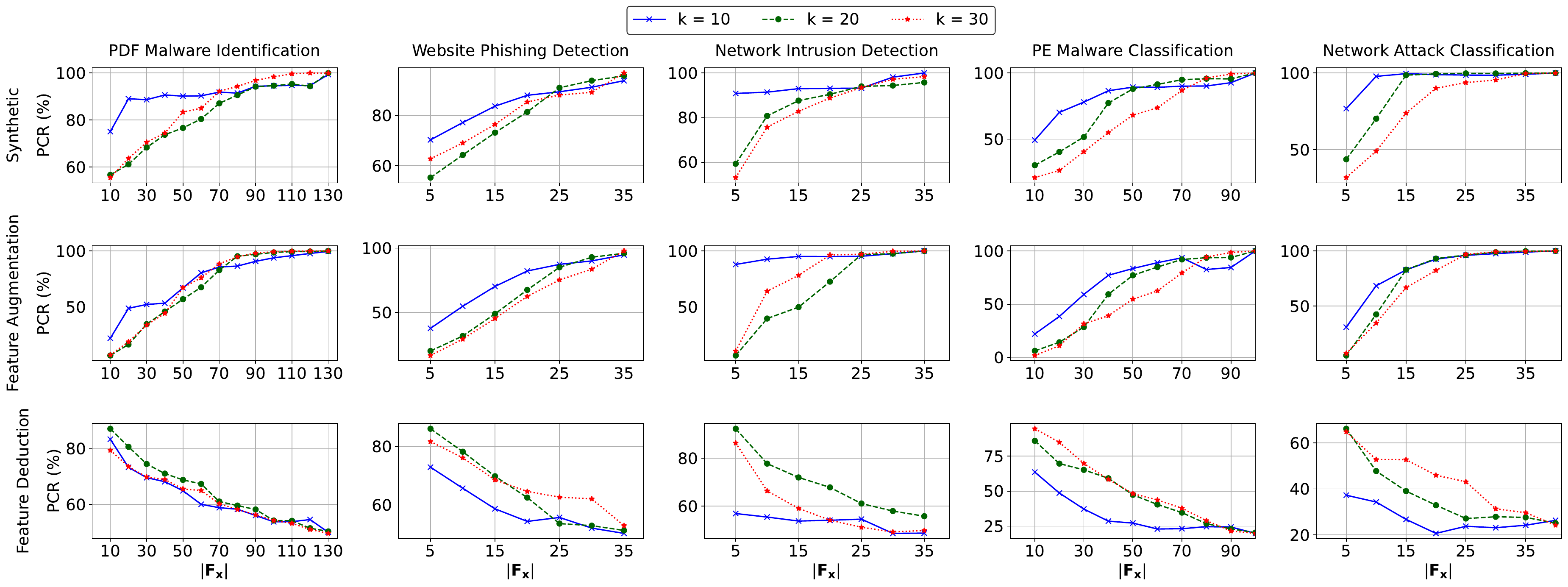}
\caption{The effect of hyper-parameter $k$ (the number of prototypes per class) on the fidelity of explanations.}
\label{fig:effect_of_k}
\end{figure*}

\shortsectionBf{Effect of Number of Prototypes.}\label{sec:effect_of_k}
To study the effect of the number of prototypes per class ($k$) on model performance and the fidelity of explanations, we train several models with identical parameters except for the value of $k$. We set $n = 3$ for this evaluation.

The performance of these models is given in Table~\ref{tab:effect_of_k}. As expected, the model is unable to capture the important prototypes for low values of $k$. Performance increases with more prototypes per class and plateaus at some value of $k$ since extra prototypes do not always encode additional useful information for classification.

\begin{table}[t!]
\centering
\caption{The effect of number of prototypes per class ($k$) on \system performance.}
\vspace{1em}
\label{tab:effect_of_k}
\resizebox{\columnwidth}{!}{
\begin{threeparttable}
\begin{tabular}{|c|c|cccc|}
\hline
\multirow{2}{*}{\textbf{Dataset}}                                               & \multirow{2}{*}{\textbf{Metric}} & \multicolumn{4}{c|}{\textbf{Number of prototypes per class}}                                                               \\ \cline{3-6} 
                                                                                &                                  & \multicolumn{1}{c|}{\textbf{k=3}} & \multicolumn{1}{c|}{\textbf{k=5}} & \multicolumn{1}{c|}{\textbf{k=20}} & \textbf{k=30} \\ \hline\hline
\multirow{3}{*}{\textbf{\begin{tabular}[c]{@{}c@{}}PDF\\ Malware\end{tabular}}}       & \textbf{Acc.}   & \multicolumn{1}{c|}{50.00\%}      & \multicolumn{1}{c|}{90.05\%}      & \multicolumn{1}{c|}{99.70\%}       & 99.80\%       \\
                                                                                      & \textbf{Pre.}   & \multicolumn{1}{c|}{50.00\%}      & \multicolumn{1}{c|}{85.47\%}      & \multicolumn{1}{c|}{100.00\%}      & 100.0\%       \\
                                                                                      & \textbf{Rec.}   & \multicolumn{1}{c|}{100.00\%}     & \multicolumn{1}{c|}{96.50\%}      & \multicolumn{1}{c|}{99.40\%}       & 99.60\%       \\
                                                                                      & \textbf{FPR}    & \multicolumn{1}{c|}{100.00\%}     & \multicolumn{1}{c|}{16.40\%}      & \multicolumn{1}{c|}{0.00\%}        & 0.00\%        \\ \hline
\multirow{3}{*}{\textbf{\begin{tabular}[c]{@{}c@{}}Website\\ Phishing\end{tabular}}}  & \textbf{Acc.}   & \multicolumn{1}{c|}{50.00\%}      & \multicolumn{1}{c|}{56.75\%}      & \multicolumn{1}{c|}{95.30\%}       & 95.15\%       \\
                                                                                      & \textbf{Pre.}   & \multicolumn{1}{c|}{-$^\dagger$}  & \multicolumn{1}{c|}{54.61\%}      & \multicolumn{1}{c|}{94.50\%}       & 93.37\%       \\
                                                                                      & \textbf{Rec.}   & \multicolumn{1}{c|}{0.00\%}       & \multicolumn{1}{c|}{80.00\%}      & \multicolumn{1}{c|}{96.20\%}       & 97.20\%       \\
                                                                                      & \textbf{FPR}    & \multicolumn{1}{c|}{0.00\%}       & \multicolumn{1}{c|}{66.50\%}      & \multicolumn{1}{c|}{5.60\%}        & 6.90\%        \\ \hline
\multirow{3}{*}{\textbf{\begin{tabular}[c]{@{}c@{}}Network\\ Intrusion\end{tabular}}} & \textbf{Acc.}   & \multicolumn{1}{c|}{62.65\%}      & \multicolumn{1}{c|}{86.05\%}      & \multicolumn{1}{c|}{99.33\%}       & 99.47\%       \\
                                                                                      & \textbf{Pre.}   & \multicolumn{1}{c|}{60.08\%}      & \multicolumn{1}{c|}{78.19\%}      & \multicolumn{1}{c|}{99.01\%}       & 99.11\%       \\
                                                                                      & \textbf{Rec.}   & \multicolumn{1}{c|}{75.40\%}      & \multicolumn{1}{c|}{100.00\%}     & \multicolumn{1}{c|}{99.65\%}       & 99.85\%       \\
                                                                                      & \textbf{FPR}    & \multicolumn{1}{c|}{50.10\%}      & \multicolumn{1}{c|}{27.90\%}      & \multicolumn{1}{c|}{1.00\%}        & 0.90\%        \\ \hline
\multirow{3}{*}{\textbf{\begin{tabular}[c]{@{}c@{}}PE\\ Malware\end{tabular}}}        & \textbf{Acc.}   & \multicolumn{1}{c|}{23.25\%}      & \multicolumn{1}{c|}{23.08\%}      & \multicolumn{1}{c|}{95.13\%}       & 95.23\%       \\
                                                                                      & \textbf{Pre.}   & \multicolumn{1}{c|}{4.651\%}      & \multicolumn{1}{c|}{4.617\%}      & \multicolumn{1}{c|}{95.63\%}       & 95.62\%       \\
                                                                                      & \textbf{Rec.}   & \multicolumn{1}{c|}{20.00\%}      & \multicolumn{1}{c|}{20.00\%}      & \multicolumn{1}{c|}{95.69\%}       & 95.80\%       \\
                                                                                      & \textbf{FPR}    & \multicolumn{1}{c|}{20.00\%}      & \multicolumn{1}{c|}{20.00\%}      & \multicolumn{1}{c|}{1.24\%}        & 1.21\%        \\ \hline
\multirow{3}{*}{\textbf{\begin{tabular}[c]{@{}c@{}}Network\\ Attack\end{tabular}}}    & \textbf{Acc.}   & \multicolumn{1}{c|}{25.00\%}      & \multicolumn{1}{c|}{99.25\%}      & \multicolumn{1}{c|}{99.83\%}       & 99.87\%       \\
                                                                                      & \textbf{Pre.}   & \multicolumn{1}{c|}{6.250\%}      & \multicolumn{1}{c|}{99.25\%}      & \multicolumn{1}{c|}{99.83\%}       & 99.88\%       \\
                                                                                      & \textbf{Rec.}   & \multicolumn{1}{c|}{25.00\%}      & \multicolumn{1}{c|}{99.25\%}      & \multicolumn{1}{c|}{99.83\%}       & 99.87\%       \\
                                                                                      & \textbf{FPR}    & \multicolumn{1}{c|}{25.00\%}      & \multicolumn{1}{c|}{0.25\%}       & \multicolumn{1}{c|}{0.056\%}       & 0.042\%       \\ \hline
\end{tabular}
\begin{tablenotes}[flushleft]
\item $^\dagger$ $-$ means no samples were classified positive; thus, precision is undefined.
\end{tablenotes}
\end{threeparttable}}
\end{table}

Using a very low number of prototypes per class ($k = 3$), \system does not have enough capacity to learn effective prototypes, leading to a drastic decrease in performance. However, at $k = 20$, the model is able to perform effectively on all three metrics. We also show the impact of $k$ on explanation fidelity, given in Figure~\ref{fig:effect_of_k}. For the PDF malware dataset, explanation fidelity does not necessarily improve with increasing $k$; in fact, it suffers slightly since more noise is built into the explanation process. The same observations apply to the phishing website and PE malware datasets, in which the fidelity is highest with $k=10$ and decreases for higher values of $k$. This effect is most prominent for the network intrusion and network attack datasets, as \system performs quite poorly with higher values of $k$. This is because the feature dimensionality is low, and learning many prototypes with very few features makes the training process result in prototypes that are unable to represent a useful embedding space.

\begin{figure*}[t!]
\centering
\includegraphics[width=\textwidth]{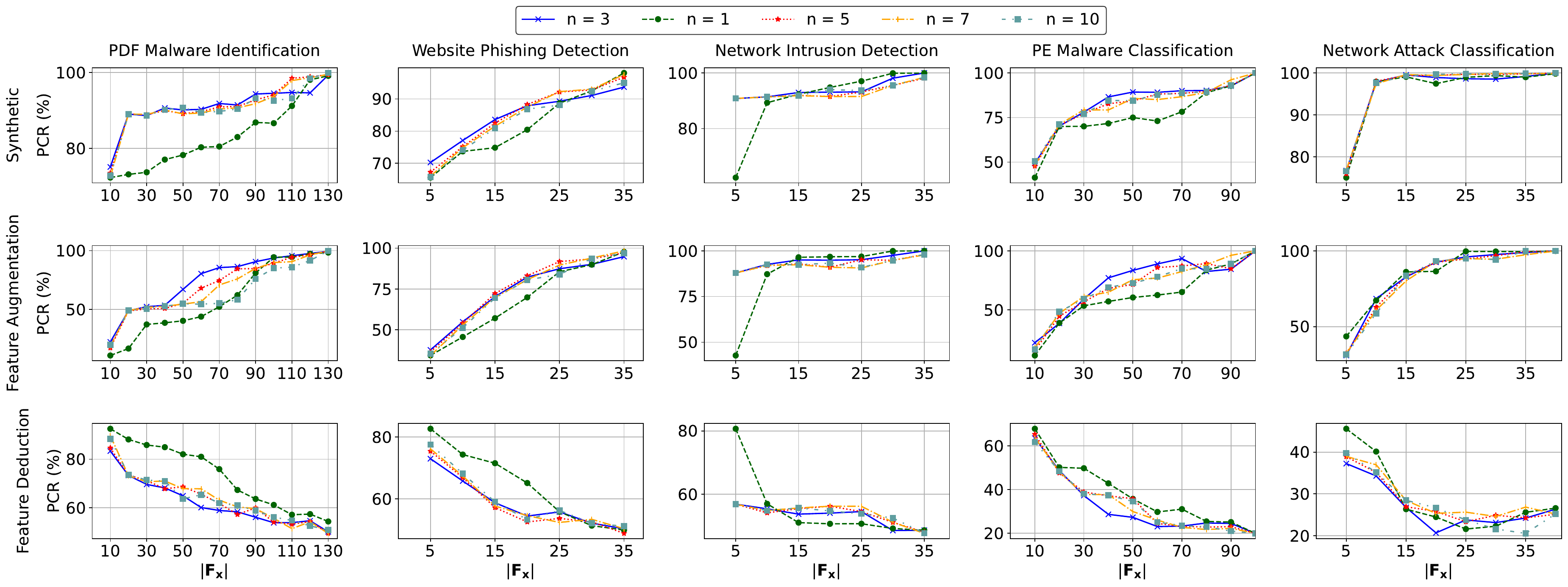}
\caption{The effect of hyper-parameter $n$ (the number of similarity scores to consider) on the fidelity of explanations.}
\label{fig:effect_of_n}
\end{figure*}

\shortsectionBf{Effect of Number of Similarity Scores.}\label{sec:effect_of_n}
The hyper-parameter $n$ controls the number of highest similarity scores to consider while generating an explanation. The value of $n$ may change the explanations generated as the importance scores are a sum of masks weighted by similarity scores. We train a model with $k = 10$ and study the effect of $n$ on explanation fidelity.

As seen for the PDF malware dataset, all three fidelity tests suffer if the value of $n$ is very low (\eg $1$). Intuitively, this is expected since a single prototype is unable to capture enough information to completely explain a sample. However, as $n$ increases, it reaches a point where all information has already been acquired, and the use of additional similarity scores and masks only adds noise to the explanation. It is important to note that the fidelity of the explanations does not suffer for high values of $n$ until very high values of $\lvert \mathbf{F}_\mathbf{{x}} \rvert$ are used. This is consistent with intuition, as the first few features are assigned very high importance scores and are always dominant contributors. When the value of $\lvert \mathbf{F}_{\mathbf{x}} \rvert$ is high, a lot of noise is injected into the tests and fidelity suffers.

For the phishing website and the PE malware datasets, it can again be seen that a single prototype is not effective in being able to generate good explanations. Moreover, the figure shows that explanation fidelity is best with $n = 3$, which shows that more prototypes capture useful information. We can again see here that the fidelity of the explanation decreases for large values of $n$ when large explanation set sizes (high value of $\lvert \mathbf{F}_\mathbf{{x}} \rvert$) are used. This is consistent with the observation for the PDF malware dataset. Interestingly, even though the feature dimensionality is low for the network intrusion and network attack datasets, the fidelity of the explanations is low for $n = 1$ at low values of $\lvert \mathbf{F}_\mathbf{{x}} \rvert$. However, the performance rapidly increases as more features are considered. As in other datasets, $n = 3$ performs the best, and fidelity does not improve much even with more prototypes being considered.

\subsection{Ablation Study}\label{sec:ablation_study}

\begin{figure*}[t!]
\centering
\includegraphics[width=\textwidth]{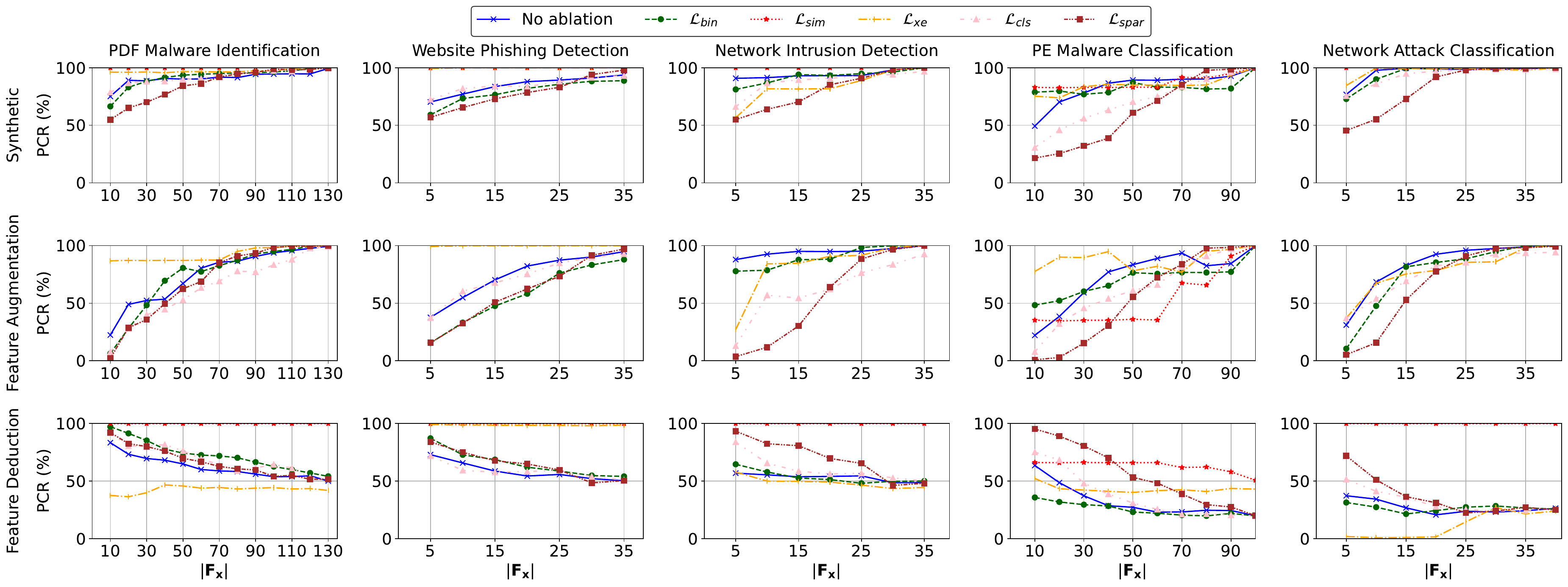}
\caption{Effect of ablating loss terms on explanation fidelity.}
\label{fig:ablation_study_fidelity}
\end{figure*}

We ablate the explanation-specific loss terms in the joint objective (Equation~\ref{eq:loss}) to evaluate their effect on explanation quality. Figure~\ref{fig:ablation_study_fidelity} shows the effect on fidelity. Removing the sparsity loss $\mathcal{L}_{spar}$, binary-mask loss $\mathcal{L}_{bin}$, or cluster loss $\mathcal{L}_{cls}$ degrades fidelity on several datasets, indicating that these terms are important for producing reliable prototype-based explanations. The feature-augmentation curves for the model without $\mathcal{L}_{sim}$ are omitted because this ablation collapses to predicting a single class, making the test inapplicable.

\begin{figure*}[t!]
\centering
\includegraphics[width=\textwidth]{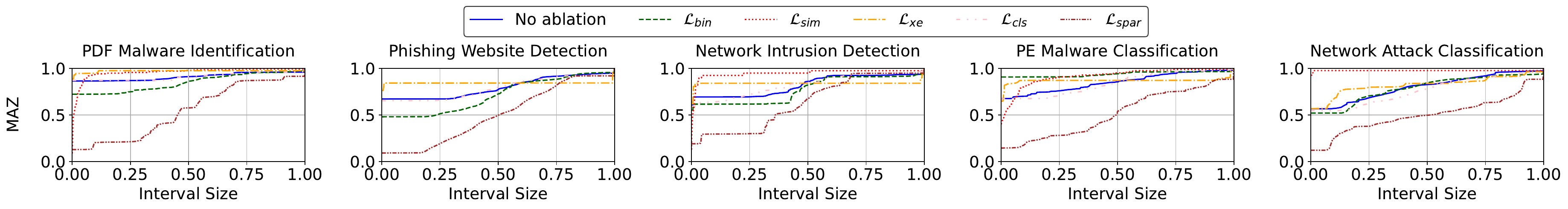}
\caption{Impact of removing each loss term on explanation sparsity.}
\label{fig:ablation_study_sparsity}
\end{figure*}

Figure~\ref{fig:ablation_study_sparsity} shows the effect on sparsity. As expected, removing $\mathcal{L}_{spar}$ reduces explanation sparsity. We also observe a collapse case on the phishing website dataset when $\mathcal{L}_{sim}$ is removed: the model learns all-zero masks, yielding zero importance scores for all samples. This behavior is consistent with the sparsity and binary-mask losses favoring zero-valued masks when no diversity term discourages identical masks. Stability is $1.0$ for all ablations because \system's explanation procedure is deterministic for a fixed model and input.

\subsection{Scalability over Feature Dimensionality}\label{sec:scalability}
We perform a scalability analysis of \system over the number of features. We use the PE malware dataset for this experiment as it has a high feature dimensionality to perform the analysis. We repeat our fidelity and sparsity experiments with $100$, $200$, and $500$ features selected by their high ANOVA F-values. The model architecture remains the same for each feature dimensionality, except for the number of prototypes per class $k$, which is set to $10$, $20$, and $50$.

\begin{table}[t!]
\centering
\caption{Classification performance of \system on the PE Malware dataset with varying feature dimensionality.}
\vspace{1em}
\label{tab:metrics_dim}
\resizebox{0.7\columnwidth}{!}{
\begin{tabular}{c|ccc|}
\cline{2-4}
\multicolumn{1}{l|}{}                    & \multicolumn{3}{c|}{\textbf{Feature Dimensionality}}                                    \\ \cline{2-4} 
\textbf{}                                & \multicolumn{1}{c|}{\textbf{100}}  & \multicolumn{1}{c|}{\textbf{200}}  & \textbf{500}  \\ \hline
\multicolumn{1}{|c|}{\textbf{Accuracy}}  & \multicolumn{1}{c|}{$94.79\%$}     & \multicolumn{1}{c|}{$95.62\%$}     & $96.36\%$     \\ \hline
\multicolumn{1}{|c|}{\textbf{Precision}} & \multicolumn{1}{c|}{$95.30$\%}     & \multicolumn{1}{c|}{$96.09$\%}     & $96.79$\%     \\ \hline
\multicolumn{1}{|c|}{\textbf{Recall}}    & \multicolumn{1}{c|}{$95.38$\%}     & \multicolumn{1}{c|}{$96.14$\%}     & $96.77$\%     \\ \hline
\multicolumn{1}{|c|}{\textbf{FPR}}    & \multicolumn{1}{c|}{$1.33$\%}      & \multicolumn{1}{c|}{$1.12$\%}      & $0.93$\%      \\ \hline
\end{tabular}}
\end{table}

\begin{figure*}[t!]
\centering
\includegraphics[width=\textwidth]{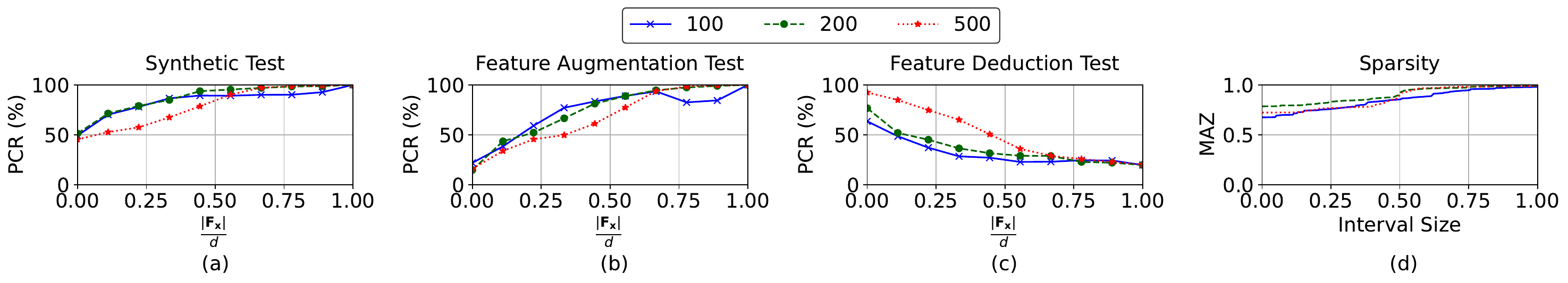}
\caption{Scalability analysis of the explanation fidelity (a, b, c) and sparsity (d) of \system. The $x$-axis for (a), (b), (c) is the fraction of features modified or removed.}
\label{fig:fidelity_dim}
\end{figure*}

The results are given in Table~\ref{tab:metrics_dim} and Figure~\ref{fig:fidelity_dim}. As expected, we notice a slight increase ($\approx 1.5\%$) in performance when the number of features increases from $100$ to $500$. Yet, explanation fidelity of \system improves with fewer features. Further, we notice that sparsity is similar for all three models. However, this is not strictly true in terms of human interpretability. This is because an explanation of $10$ features for the model with $100$ features is considered to be as sparse as an explanation of $50$ features for the model with $500$ features. Yet, $10$ features are far more interpretable by humans than $50$ features.

Based on the above observations, we recommend using dimensionality reduction techniques before training \system. By doing this, the sparsity of explanations improves with little effect on the fidelity or classification accuracy.

\subsection{Case Study}\label{sec:case_study}
We provide a qualitative case study to illustrate how \system's explanations can be inspected by an analyst. We focus on network intrusion detection because its flow-level features are relatively interpretable, but the same analysis can be applied to the other datasets. Our goal is to show that \system highlights features that are consistent with class-discriminative patterns in the data.

In the network intrusion dataset, we find that minimum time to live (\texttt{MIN\_TTL}) is a strong indicator of whether a flow is benign or malicious. This observation is dataset-specific: \texttt{MIN\_TTL} is not universally indicative of malicious traffic, but in this dataset its distribution differs substantially between benign and attack flows. Most benign flows have \texttt{MIN\_TTL}$=31$, whereas malicious flows often have different, typically higher, values. Since TTL values reflect network path and host-configuration characteristics, this feature helps distinguish benign and attack traffic in this dataset. Thus, for a benign flow with low \texttt{MIN\_TTL}, we expect \system to classify the sample as benign and assign high importance to \texttt{MIN\_TTL}; similarly, for a malicious flow with high \texttt{MIN\_TTL}, we expect \system to identify \texttt{MIN\_TTL} as evidence for the malicious prediction.

\begin{table}[t!]
\centering
\caption{A case study of the network intrusion dataset. The table shows examples along with their ground truth (\textbf{GT}), the value of \texttt{MIN\_TTL}, prediction (\textbf{Pred}), and the top-ranked important features identified by \system. The most important features are ordered by row--the first row contains features with the highest scores.}
\vspace{1em}
\label{tab:cs_netflow}
\setlength{\tabcolsep}{3pt}
\resizebox{\columnwidth}{!}{
\begin{threeparttable}
\begin{tabular}{|c|c|c|l|}
\hline 
\textbf{GT} & \textbf{\texttt{MIN\_TTL}} & \textbf{Pred} & \multicolumn{1}{|c|}{\textbf{Explanation}} \\ \hline\hline
\multirow{3}{*}{Benign}     & \multirow{3}{*}{31}     & \multirow{3}{*}{Benign}    & \texttt{DNS\_QUERY\_TYPE}$ = 1$, \texttt{TCP\_FLAGS}$ = 0$, \texttt{MIN\_TTL}$ = 31.0$ \\ 
&      &     & \texttt{MIN\_IP\_PKT\_LEN}$ = 73$ \\ 
&      &     & \texttt{DNS\_TTL\_ANSWER}$ = 60$ \\ \hline

\multirow{3}{*}{Malicious}  & \multirow{3}{*}{254}    & \multirow{3}{*}{Malicious} & \texttt{NUM\_PKTS\_128\_TO\_256\_BYTES}$ = 0$, \texttt{IN\_PKTS}$ = 2$ \\ 
  &     &  & \texttt{MIN\_TTL}$ = 254$ \\ 
  &     &  & \texttt{LONGEST\_FLOW\_PKT}$ = 100$ \\ \hline
\end{tabular}
\end{threeparttable}}
\end{table}

Table~\ref{tab:cs_netflow} shows one true negative and one true positive example. In both cases, \system ranks \texttt{MIN\_TTL} among the top-$3$ most important features. For the true negative sample, \texttt{MIN\_TTL}$=31$, and \system correctly predicts the flow as benign. For the true positive sample, \texttt{MIN\_TTL}$=254$, and \system correctly predicts the flow as malicious. These examples show that \system's explanations align with a feature that separates benign and malicious flows in the dataset.

We also compare these explanations with LIME and SHAP using an MLP target model that correctly classifies the same two samples. For the benign sample, LIME ranks \texttt{MIN\_TTL} as the $10$th most important feature and SHAP ranks it $8$th. For the malicious sample, LIME ranks \texttt{MIN\_TTL} $31$st, and SHAP ranks it $10$th. In contrast, \system ranks \texttt{MIN\_TTL} in the top $3$ for both samples, making the class-discriminative feature more prominent in the explanation.
\section{Limitations and Discussion}
\label{sec:discussion}

\shortsectionBf{Robustness to Adversarial Perturbations.} 
In the context of XAI, adversarial perturbations aim to change ($1$) the model prediction, or ($2$) the generated explanations while retaining the predicted label~\cite{dombrowski_2019, slack_2020}. Formally, if $f$ is the classifier and $e$ is the XAI method that operates on $f$ and a data point $x$, the goal of such attacks is to find a perturbation $\delta$ such that $f(x + \delta) \neq f(x)$ or $e(f, x + \delta) \neq e(f, x)$.

Prior work shows that XAI methods are vulnerable to white-box attacks that can produce arbitrary explanations~\cite{dombrowski_2019} or labels~\cite{zhang_2020}. Recent work~\cite{slack_2020} has shown that post-hoc approximation methods, \eg LIME and SHAP, are vulnerable to attacks that create arbitrary explanations that do not reflect the biases in the original model. 

We evaluate the robustness of \system in the presence of adversarial perturbations that aim to achieve each of the above objectives on the Website Phishing dataset. We leverage FGSM~\cite{fgsm} and PGD~\cite{pgd} to evaluate the effect of perturbations on predictions of \system and ProtoPNet. The attack success rates are presented in Table~\ref{tab:robustness_pred}. Both \system and ProtoPNet are affected similarly with increasing ASR as attack budget increases.

\begin{table}[t!]
    \centering
    \caption{Results of robustness analysis of \system and ProtoPNet. FGSM and PGD aim to change the predicted label with varying levels of adversarial budget $\epsilon$. Each cell displays the attack success rate (ASR) with the corresponding setup.}
    \vspace{1em}
    \label{tab:robustness_pred}
    \setlength{\tabcolsep}{0.6em}
    \resizebox{\columnwidth}{!}{
    \begin{threeparttable}
    \begin{tabular}{|c|ccc|ccc|}
    \hline
                           & \multicolumn{3}{c|}{\textbf{FGSM}}                              & \multicolumn{3}{c|}{\textbf{PGD}}                               \\ \hline
                           & \multicolumn{1}{c|}{$\epsilon=0.01$}  & \multicolumn{1}{c|}{$\epsilon=0.1$}   & $\epsilon=0.3$   & \multicolumn{1}{c|}{$\epsilon=0.01$}  & \multicolumn{1}{c|}{$\epsilon=0.1$}   & $\epsilon=0.3$   \\ \hline
    \textbf{\system} & \multicolumn{1}{c|}{5.75\%} & \multicolumn{1}{c|}{14.45\%} & 46.95\% & \multicolumn{1}{c|}{5.75\%} & \multicolumn{1}{c|}{14.70\%} & 52.15\% \\ \hline
    \textbf{ProtoPNet}     & \multicolumn{1}{c|}{6.55\%} & \multicolumn{1}{c|}{28.35\%} & 41.45\% & \multicolumn{1}{c|}{6.80\%} & \multicolumn{1}{c|}{38.15\%} & 84.15\% \\ \hline
    \end{tabular}
    \end{threeparttable}}
\end{table}

We also modify an existing targeted attack~\cite{dombrowski_2019} into an untargeted attack to study how adversarial perturbations affect the explanations generated by \system while preserving the predicted label. We evaluate the same attack on LIME and report the results in Table~\ref{tab:robustness_expl}. We use two metrics: attack success rate (ASR), defined here as the proportion of samples whose predicted label remains unchanged after perturbation, and cosine similarity between the original explanation and the explanation for the perturbed sample. A higher ASR indicates that the attack successfully preserves the model prediction, while a lower cosine similarity indicates a larger change in the explanation.

The attack preserves the predicted label for most samples for both methods. However, \system's explanations change less than LIME's under this attack, as reflected by the higher cosine similarity. These results suggest that \system provides better explanation stability than LIME under this specific perturbation setting. However, they do not establish general robustness against adaptive explanation attacks, which remains an important direction for future work.

\begin{table}[t!]
\centering
\small
\caption{Results of an existing attack~\cite{dombrowski_2019} on \system and LIME. The attack attempts to change the explanation arbitrarily without affecting the prediction.}
\vspace{1em}
\label{tab:robustness_expl}
\setlength{\tabcolsep}{3pt}
\begin{threeparttable}
\begin{tabular}{|c|c|c|}
\hline
                & \textbf{\begin{tabular}[c]{@{}c@{}}Attack Success Rate\end{tabular}} & \textbf{\begin{tabular}[c]{@{}c@{}}Cosine Similarity\end{tabular}} \\ \hline
\textbf{\system} & 99.55\%                                                                & 0.754                                                                \\ \hline
\textbf{LIME}   & 100.0\%                                                                & 0.131                                                                \\ \hline
\end{tabular}
\end{threeparttable}
\end{table}

\shortsectionBf{Failure Analysis.}
We examine two network intrusion examples where \system misclassifies the sample. These examples illustrate how \system's explanations can help diagnose model errors by revealing when the prediction relies on features whose values are atypical or inconsistent with the expected class pattern.
 
Table~\ref{tab:cs_netflow_failure} shows one false negative and one false positive example. In the false negative case, \system misclassifies a malicious flow as benign because it assigns high importance to the low value of \texttt{MIN\_TTL}, which is typically associated with benign traffic in this dataset. Further inspection shows that this sample is anomalous, with most feature values equal to $0$. In the false positive case, \system predicts a benign flow as malicious because \texttt{MIN\_TTL}$=254$, a value more commonly associated with malicious flows in our case study (see Section~\ref{sec:case_study}). These examples show that \system's explanations can help analysts identify when errors arise from atypical feature patterns rather than opaque model behavior.

\begin{table}[t!]
\centering
\caption{Examples of failure cases of \system on the network intrusion dataset.}
\vspace{1em}
\label{tab:cs_netflow_failure}
\setlength{\tabcolsep}{3pt}
\resizebox{\columnwidth}{!}{
\begin{threeparttable}
\begin{tabular}{|c|c|c|l|}
\hline 
\textbf{GT} & \textbf{\texttt{MIN\_TTL}} & \textbf{Pred} & \multicolumn{1}{|c|}{\textbf{Explanation}} \\ \hline\hline
\multirow{2}{*}{Malicious}     & \multirow{2}{*}{31}     & \multirow{2}{*}{Benign}    & \texttt{DNS\_QUERY\_TYPE}$ = 0$, \texttt{TCP\_FLAGS}$ = 0$, \texttt{MIN\_TTL}$ = 0$ \\ 
     &      &     & \texttt{RETRANSMITTED\_OUT\_BYTES}$ = 0$, \texttt{DNS\_TTL\_ANSWER}$ = 0$ \\ \hline
\multirow{3}{*}{Benign}  & \multirow{3}{*}{254}    & \multirow{3}{*}{Malicious} & \texttt{NUM\_PKTS\_128\_TO\_256\_BYTES}$ = 0$, \texttt{IN\_PKTS}$ = 2$ \\ 
  &     &  & \texttt{MIN\_TTL}$ = 254$, \texttt{DST\_TO\_SRC\_SECOND\_BYTES}$ = 0$ \\ 
  &     &  & \texttt{LONGEST\_FLOW\_PKT}$ = 44$ \\ \hline
\end{tabular}
\end{threeparttable}}
\end{table}

\shortsectionBf{Guidelines for Training.}
The goal when training \system is to balance predictive performance with explanation quality. A self-explainable model with poor classification performance is not useful, even if its explanations are sparse or stable; conversely, an accurate model may not produce the most faithful or interpretable explanations. In practice, we first tune the architecture, learning rate, batch size, embedding dimension, and number of prototypes per class to achieve strong validation accuracy. We then tune the loss coefficients to improve explanation fidelity and sparsity while maintaining comparable classification performance.

The explanation-specific losses should be adjusted jointly. Increasing the sparsity loss encourages concise explanations, but overly strong sparsity can remove informative features or lead to degenerate masks. The binary-mask loss should be large enough to produce interpretable near-binary masks, while the mask-similarity loss helps prevent different prototypes from collapsing to the same feature subset. The cluster loss should be tuned to maintain meaningful prototype alignment without sacrificing classification performance. We therefore select the final model based on validation accuracy, fidelity, and sparsity, rather than accuracy alone.
\section{Conclusion}
We introduce \system, a prototype-based self-XAI architecture that jointly performs classification and generates feature-importance scores as explanations. \system learns class-specific prototypes and prototype-specific masks, using prototype similarities for prediction and explanation. Across five security datasets, \system achieves competitive predictive performance while producing explanations that are faithful, sparse and perfectly stable. These results show that \system is a significant advancement towards interpretable security models.

\section*{Acknowledgments}

This material is based upon work supported by the National Science Foundation (NSF) under grant no. 2229876 and is supported in part by funds provided by the NSF, by the Department of Homeland Security, and by IBM. Any opinions, findings, and conclusions or recommendations expressed in this material are those of the author(s) and do not necessarily reflect the views of the NSF or its federal agency and industry partners.

\bibliographystyle{plainurl}
\bibliography{references}

\appendix
\section{Details of Datasets}\label{app:dataset_details}

Here we present details of the datasets we use for evaluation.

\shortsectionBf{PDF Malware Identification.} This dataset includes features (\eg number of characters in the title) extracted from PDF files labeled malware or benign~\cite{mimicus, pdfmalware}. We use $135$ features extracted from $\sim 5$K benign and $5$K malicious (malware) PDF files, following previous work~\cite{lemna}. 

\shortsectionBf{Phishing Website Detection.} We use $38$ features (\eg URL length, presence of specific keywords associated with phishing) from a phishing dataset~\cite{phishing} that contains features of a website labeled either as a phishing website or a benign website. $5$K phishing web page URLs are collected from PhishTank and OpenPhish, and $5$K legitimate web page URLs are from Alexa and the Common Crawl5 archive.

\shortsectionBf{Network Intrusion Detection.} We use $39$ flow-based features (\eg packet counts, byte counts, flow duration) from the Netflow dataset~\cite{sarhan_2022}, an extension of the popular dataset UNSW-NB15~\cite{moustafa_2015, moustafa_2016, moustafa_2017, moustafa_2019, sarhan_2021}. We randomly sample $10$K benign and $10$K attack samples for our experiments.

\shortsectionBf{PE Malware Classification.} The BODMAS dataset~\cite{bodmas} includes $2{,}381$ features on benign and $581$ malware families extracted from program executables. We use all samples from malware families that have at least $3{,}000$ samples, which are $5$ classes. We then remove constant features and select $100$ features with the highest ANOVA F-value scores for our experiments.

\shortsectionBf{Network Attack Classification.} The NSL-KDD dataset~\cite{tavallaee_2009_detailed} is a widely used dataset that has flow-based features with temporal relationships. The labels are either benign or the network attack type. We preprocess the dataset and select all attacks that have at least $3{,}000$ samples, \ie neptune, satan, ipsweep, and smurf attacks.

\section{Hyperparameters of \system}\label{app:hyperparameters}
The hyperparameters of \system's training on each dataset are given in Table~\ref{tab:hyper_parameters}. The feature encoder is an MLP for the PDF malware, phishing website, network intrusion, and PE malware datasets and an LSTM for the network attack dataset. This is because the network attack dataset has temporal relationships between features, which are better modeled by an LSTM.

\begin{table}[t!]
\centering
\caption{Hyper-parameters and network architecture details obtained by training \system on each dataset. $\Lambda$ is the vector of coefficients in the same order as Equation~\ref{eq:loss}.}
\vspace{1em}
\label{tab:hyper_parameters}
\setlength{\tabcolsep}{3pt}
\resizebox{\columnwidth}{!}{
\begin{tabular}{|c|c|c|c|c|c|c|c|}
\hline
\textbf{Dataset} & \textbf{\begin{tabular}[c]{@{}c@{}}Learning\\ Rate\end{tabular}} & \textbf{\begin{tabular}[c]{@{}c@{}}\# Neurons\\ in $\mathcal{M}$ \end{tabular}}& \textbf{\begin{tabular}[c]{@{}c@{}}\# Neurons\\ in $f$ \end{tabular}} & \textbf{\begin{tabular}[c]{@{}c@{}}Embedding\\ Dimension\end{tabular}} & \textbf{\begin{tabular}[c]{@{}c@{}}$k$\end{tabular}} & \textbf{\begin{tabular}[c]{@{}c@{}}Batch\\ Size\end{tabular}} & $\Lambda$ \\ \hline\hline
\begin{tabular}[c]{@{}c@{}} PDF\\ Malware \end{tabular}      & $1\mathrm{e}{-3}$                                                             & \begin{tabular}[c]{@{}c@{}} 512, 1024,\\ 2048 \end{tabular}            & 256, 128, 64             & 128                                                                    & 10   &     256               &     \begin{tabular}[c]{@{}c@{}} 1.0, 3.5, 0.8,\\ 0.5, 0.5 \end{tabular}    \\ \hline
\begin{tabular}[c]{@{}c@{}} Website\\ Phishing \end{tabular}          & $5\mathrm{e}{-3}$                                                             & \begin{tabular}[c]{@{}c@{}} 256, 512, \\ 1024 \end{tabular}       & 256, 128, 64             & 128                                                                    & 10   &       256           &    \begin{tabular}[c]{@{}c@{}} 3.5, 5.0, 0.5,\\ 0.5, 0.5 \end{tabular}   \\ \hline
\begin{tabular}[c]{@{}c@{}} Network\\ Intrusion \end{tabular}        & $1\mathrm{e}{-3}$                                                             & \begin{tabular}[c]{@{}c@{}} 512, 1024,\\ 2048 \end{tabular}           & 256, 128, 64             & 128                                                                    & 10   &     256                &    \begin{tabular}[c]{@{}c@{}} 1.0, 5.0, 0.8,\\ 0.5, 0.5 \end{tabular}  \\ \hline
\begin{tabular}[c]{@{}c@{}} PE\\ Malware \end{tabular}      & $1\mathrm{e}{-3}$                                                             & \begin{tabular}[c]{@{}c@{}} 256, 512, \\ 1024 \end{tabular}      & 256, 128, 64             & 128                                                                    & 10   &     256                          &    \begin{tabular}[c]{@{}c@{}} 1.0, 15.0, 0.8,\\ 0.5, 0.5 \end{tabular}    \\ \hline
\begin{tabular}[c]{@{}c@{}} Network\\ Attack \end{tabular}      & $1\mathrm{e}{-3}$                                                             & \begin{tabular}[c]{@{}c@{}} 256, 512, \\ 1024 \end{tabular}      & 4             & 4                                                                    & 10   &     256                                 &  \begin{tabular}[c]{@{}c@{}} 1.0, 15.0, 0.8,\\ 0.5, 0.5 \end{tabular}    \\ \hline
\end{tabular}}
\end{table}

\end{document}